\documentclass[11pt]{article}

\usepackage[letterpaper, margin=1in]{geometry}
 \usepackage{setspace}
\usepackage[hidelinks]{hyperref}
\usepackage{amsmath, amsthm, amsfonts, 
thmtools, amssymb,tikz,cleveref}
\usepackage{bbm}
\usepackage[mathscr]{euscript}
\numberwithin{equation}{section}
\usepackage{comment}
\usepackage{authblk}

\DeclareMathOperator{\var}{Var}
\usepackage{algpseudocode}

\usepackage{algorithm}

\newtheorem{thm}{Theorem}[section]
\newtheorem{prop}[thm]{Proposition}
\newtheorem{lem}[thm]{Lemma}
\newtheorem{cor}[thm]{Corollary}

\theoremstyle{remark}
\newtheorem{remark}[thm]{Remark}
\theoremstyle{definition}
 
\newtheorem{definition}[thm]{Definition}

\newtheorem{alg}[thm]{Algorithm}

\newcommand{\E}{\mathbb{E}}
\renewcommand{\P}{\mathbb{P}}

\newcommand{\bbR}{\mathbb{R}}

\renewcommand{\phi}{\varphi}

\newcommand{\one}{\mathbbm{1}}
\newcommand{\bs}{\boldsymbol}

\newcommand{\be}{\begin{equation}}
\newcommand{\ee}{\end{equation}}
\newcommand{\bex}{\begin{equation*}}
\newcommand{\eex}{\end{equation*}}
\newcommand{\unn}[2]{[\![#1,#2]\!]}

\renewcommand{\tilde}{\widetilde}
\newcommand{\eps}{\varepsilon}

\title{Fast computation of exact confidence intervals for randomized experiments with binary outcomes}
\author[a]{P.\ M.\ Aronow}
\author[b]{Haoge Chang}
\author[c]{Patrick Lopatto\footnote{Corresponding author. E-mail address: \texttt{lopatto@unc.edu}}}
\affil[a]{\small{Departments of Statistics and Data Science, Political Science, Biostatistics and Economics, Yale University, 115 Prospect Street, New Haven, CT 06520}}
\affil[b]{Department of Economics, Columbia University, 1022 International Affairs Building, 420 West 118th Street, New York, NY 10027}
\affil[c]{Department of Statistics and Operations Research, UNC--Chapel Hill, 150 E.\ Cameron Ave, Chapel Hill, NC 27599}
\begin{document}
\maketitle
\thispagestyle{empty}
\begin{abstract}
Given a randomized experiment with binary outcomes, exact confidence intervals for the average causal effect of the treatment can be computed 
through a series of permutation tests. This approach requires minimal assumptions and is valid for all sample sizes, as it does not rely on large-sample approximations such as those implied by the central limit theorem. We show that these confidence intervals can be found in $O(n \log n)$  permutation tests in the case of balanced designs, where the treatment and control groups have equal sizes, and $O(n^2)$ permutation tests in the general case. 
Prior to this work, the most efficient known constructions
required $O(n^2)$ such tests in the balanced case  [Li and Ding, 2016], and $O(n^4)$ tests in the general case [Rigdon and Hudgens, 2015]. 
Our results thus facilitate exact inference as a viable option for randomized experiments far larger than those accessible by previous methods. We also generalize our construction to produce confidence intervals for other causal estimands, including the relative risk ratio and odds ratio, yielding similar computational gains. 
\end{abstract}

\newpage 

\setcounter{page}{1}

\section{Introduction}\label{s:introduction}

There is a vast literature on causal inference for randomized experiments with binary outcomes (see, for example, the discussion and references in \cite{imbens2015causal}). This setting covers many important cases, such as trials testing the effect of a new drug on the occurrence of a medical condition (e.g., stroke or heart attack). By the end of the study, each subject either has, or has not, been diagnosed with the condition. Other examples include experiments comparing online advertisements  that subjects either click or ignore, and studies of voting incentives, where subjects either vote or abstain.

The study of causal inference for binary outcomes  leads to 
algorithmic questions, which are the focus of this article. To contextualize our results, we first review the relevant statistical background. We then describe previous work on algorithms for inference on randomized experiments with binary outcomes and present our contributions. 

\subsection{Statistical Background}

Many traditional approaches to constructing confidence intervals for experiments with binary outcomes are based on a binomial model for the observed subject outcomes. These include Wald  intervals \cite[Chapter 10]{wasserman2004all}, which are based on a normal approximation valid in large samples, as well as other interval estimators \cite{santner1980small}.
However, the assumptions underlying
the binomial model are problematic in typical experimental designs. 
This was noted by Robins in \cite{robins1988confidence}, whose discussion we now  summarize.

Consider an experiment with $n$ subjects, where $m$ subjects are randomly assigned to treatment and the remaining $n-m$ are assigned to control, and the random assignment is such that all possible configurations are equally likely. 
Robins identifies two modeling assumptions under which binomial confidence intervals will cover the true average causal effect at the nominal rate, where the notion of ``true average causal effect'' is understood differently for each model. 
\begin{enumerate}

\item\label{assume1} 
If a subject is exposed to treatment, their outcome may be modeled as a Bernoulli random variable, and the treatment outcomes across subjects are independent with common mean $p_1$. Similarly, the control outcomes are independent Bernoulli random variables with common mean $p_2$, which are also independent from the treatment outcomes. Define the average causal effect of treatment as $p_1 - p_2$. This quantity is defined using only the experimental subjects. 

\item\label{assume2} Suppose that the subjects in the study are drawn uniformly at random from 
some near-infinite ``superpopulation." Let $p_1$ be the proportion of subjects in the superpopulation whose outcome would be $1$ under treatment, and similarly let $p_2$ be the proportion whose outcome would be $1$ under control. Again define the average causal effect of treatment as $p_1 - p_2$. While notationally similar to the previous model, this estimand is defined using the entire superpopulation, not only the experimental subjects.

\end{enumerate}
Robins notes that Assumption~\eqref{assume1} is untenable in most contexts, since it does not account for between-subject variation. For example, disease risk may vary among subjects in a clinical trial according to pre-existing medical conditions and demographic characteristics. Further, Assumption~\eqref{assume2} is usually false, since typical subject recruitment strategies do not result in a uniform sample from a well-defined superpopulation. Consider, for example, a clinical trial where subjects are recruited through newspaper advertisements. It is generally implausible that all members of the target population will read and respond to such advertisements at identical rates, resulting in a non-uniform sample. 
Further, when both Assumption~\eqref{assume1} and Assumption~\eqref{assume2} fail, it is unclear what causal parameter the binomial confidence intervals are supposed to estimate, and why these intervals  should cover it at the nominal rate.

Given the deficiencies of these modeling assumptions, we instead adopt a model that treats the potential outcomes of the subjects as fixed. In contrast to Assumption~\eqref{assume1}, the statistical randomness in our setup comes only from the assignment of subjects to treatment and control. And unlike the model in Assumption~\eqref{assume2}, we target an estimand defined using only the subjects in the experiment, without referring to a superpopulation. This perspective originated in work of Neyman and has subsequently been developed by many researchers (see \cite[Chapter 6]{imbens2015causal} for details).


Suppose we have a group of $n$ subjects, and label them arbitrarily from $1$ to $n$.
Our fundamental assumption is that the observed outcome for subject $i$ depends only on the index $i$ and whether that subject is assigned to treatment or control. This is known  as the \emph{stable unit treatment value assumption}, and abbreviated SUTVA. In particular, SUTVA rules out between-subject interaction effects. We let $\bs y_i = (y_i(1), y_i(0))$ with $y_i(0),y_i(1) \in \{0,1 \}$ denote the \emph{potential outcomes} for the $i$-th subject, where $y_i(1)$ is the outcome that would be observed if the subject were assigned to treatment, and $y_i(0)$ is the outcome that would be observed if the subject were assigned to control. 
In the experiment, only one of these potential outcomes is observed; the other remains unknown. 
We aim to estimate the \emph{sample average treatment effect}, defined as 
\be\label{sate}
\tau(\bs y) =  \frac{1}{n} \sum_{j=1}^n \big( y_j(1) - y_j(0) \big).
\ee

The quantity $\tau(\bs y)$ is defined using only the subjects in the experiment, and may not be informative for a different population.
Such a generalization would require additional assumptions. However, precise estimates of $\tau(\bs y)$ are still broadly useful for studying causal claims. For example, if a company asserts that their latest drug greatly reduces the chance of stroke in elderly people, and a study of the average causal effect in a group of such people produces a precise estimate centered at $0$, it is reasonable to question the company's claim.

We retain the previous experimental design, where $m$ subjects are assigned to treatment, the remainder are assigned to control, and all such assignments are equally likely.\footnote{Note that under this design and our assumptions, the number of observed $1$ outcomes is not in general binomially distributed.} 
Under this design, Robins gives a confidence interval for $\tau(\bs y)$ that relies on a large-sample normal approximation, which is explicitly justified by randomization. 
While this interval is guaranteed to cover $\tau(\bs y)$ at its nominal rate in the limit as $n$ goes to infinity, the literature does not offer a  quantitative estimate for its accuracy with finite $n$.  
We also note that other asymptotically valid confidence intervals have been derived in the fixed potential outcomes framework  under the assumption that the covariance matrix of the potential outcomes converges to a deterministic value as $n$ goes to infinity; see \cite[Theorem 12.2.4]{lehmann2022testing} and \cite{li2017general}. Again, coverage is guaranteed only in the large $n$ limit.\footnote{Under a binomial outcome model, the coverage probability of a standard Wald interval exhibits erratic behavior in finite samples and may lead to severe undercoverage. See, for example, \cite{brown2001interval}. Similar behavior can be observed in the fixed potential outcomes setting as well. See, for example, the first panel of Figure 1 in \cite{rigdon2015randomization}.} 
\subsection{Permutation Intervals}
We now explain how the construction of confidence intervals for the causal effect in a randomized experiment with a binary outcome 
may be cast as a computational problem. Specifically, we seek confidence intervals that contain $\tau(\bs y)$ with probability at least $1-\alpha$  regardless of the sample size or potential outcomes.

Define a random vector $\bs Z \in \{0,1\}^n$ by letting $Z_i = 0$ if subject $i$ is assigned to control, and $Z_i= 1$ if subject $i$ is assigned to treatment. Let $\bs Y$ be the vector of observed outcomes, where
\begin{equation}
Y_j = Z_j y_j(1) + (1 -  Z_j) y_j(0)
\end{equation}
represents the outcome that the experimenter observes for subject $j$ after the experiment is completed. Using the known vectors $\bs Z$ and $\bs Y$, the unknown parameter $\tau (\bs y)$ may be estimated using the \emph{Neyman estimator} $ T(\bs Y, \bs Z)$, defined by
\be
T(\bs Y, \bs Z) = \sum_{j=1}^n  \frac{Z_j Y_j}{m}  - \sum_{j=1}^n \frac{(1-Z_j) Y_j}{n-m}.
\ee
An elementary calculation shows that $\E [ T(\bs Y, \bs Z) ] = \tau(\bs y)$, where the expectation is taken over the variable $\bs Z$. Hence $T(\bs Y, \bs Z)$ is an unbiased estimator for $\tau(\bs y)$. 

As a preliminary step toward constructing a confidence interval for $\tau(\bs y)$, we consider the probability that some possible configuration $\bs w$ of the  potential outcomes could produce data at least as extreme as the observed data $(\bs Y, \bs Z)$. To make this notion precise, we consider an arbitrary length $n$ vector $\bs w$ of elements $\bs w_i$ such that $w_i(0),w_i(1) \in \{0,1\}$, and compute the $p$-value given by
\begin{equation}\label{intropvalue}
p(\bs w, \bs Y , \bs Z) = \P\left( \big| T(\bs{\tilde Y}, \bs{ \tilde Z} )  - \tau(\bs w) \big| \ge \big| T(\bs Y, \bs Z )  - \tau(\bs w) \big| \right).
\end{equation}
Here $\bs{ \tilde Z}$ is independent from $\bs Z$ and has an identical distribution, and $\bs {\tilde Y}$ is the vector of observed outcomes in a hypothetical experiment with potential outcomes $\bs w$ and randomization $\bs{ \tilde Z}$, so that $\tilde Y_j = \tilde Z_j w_j(1) + (1 -  \tilde Z_j) w_j(0)$. Calculating the $p$-value \eqref{intropvalue} is known as performing a \emph{permutation test}, since the probability may be computed by averaging over all possible randomizations $\bs{ \tilde Z}$.

Recall that $\E [ T(\bs {\tilde Y}, \bs {\tilde Z}) ] = \tau(\bs w)$. Then, informally speaking, the $p$-value measures how atypical (far from the center of the distribution) the observed data $T(\bs Y, \bs Z )$ is, under the assumption that $\bs w$ is the true set of potential outcomes for the subjects in the experiment. The permutation test \eqref{intropvalue} was essentially known to Fisher in the 1930s \cite[Chapter 5]{imbens2015causal}, and can be used to test the hypothesis that the true potential outcomes $\bs y$ are equal to some specific $\bs w$ (by rejecting this hypothesis if $p(\bs w, \bs Y , \bs Z)$ is sufficiently small). 

Recently, it was realized by Rigdon and Hudgens that exact confidence intervals for $\tau(\bs y)$ could be constructed from a series of these permutation tests  in the following way \cite{rigdon2015randomization}. Given observed data $(\bs Y, \bs Z)$, consider the set of all potential outcome vectors $\bs w$ that could produce $(\bs Y, \bs Z)$.\footnote{This already rules out many vectors $\bs w$. For example, if $n=2$ and we randomize into equal groups of size $1$, and the observed data yields $T(\bs Y, \bs Z) = 1$, then the configuration $\bs w_1 = \bs w_2 = (0,0)$ is not possible.}
(The binary outcome assumption ensures that this set is finite.) Fix $\alpha \in (0,1)$, and declare such a $\bs w$ \emph{compatible} if $p(\bs w, \bs Y , \bs Z) \ge \alpha$, and \emph{incompatible} otherwise. Let $L_\alpha( \bs Y, \bs Z)$ equal the smallest value of $\tau(\bs w)$ where $\bs w$ ranges over all compatible $\bs w$, and let $U_\alpha(\bs Y, \bs Z)$ equal the largest value of $\tau(\bs w)$ where $\bs w$ ranges over all compatible $\bs w$. Set $\mathcal I_\alpha  = [ L_\alpha(\bs Y, \bs Z), U_\alpha(\bs Y, \bs Z) ]$. Then it is straightforward to show that $\mathcal I_\alpha$ contains $\tau(\bs y)$ with probability at least $1-\alpha$, regardless of the value of $\bs y$.\footnote{See \Cref{l:containsobserved} in \Cref{s:basic}, below.} Further, Rigdon and Hudgens demonstrated through extensive simulation evidence that the intervals $\mathcal I_\alpha$ compare favorably to the intervals produced by large-sample normal approximations, in the sense that they have lengths similar to the large-sample intervals 
for parameter values such that the approximations have the correct coverage rate, and cover at the nominal rate even when the approximations do not.

\subsection{Algorithmic Perspective and Main Results}
Constructing $\mathcal I_\alpha$ through permutation tests eliminates any reliance on uncontrolled large-sample approximations, but has the drawback of requiring significant computational resources. While a naive construction of $\mathcal I_\alpha$ might proceed by directly computing \eqref{intropvalue} for all possible $\bs w$ to find $L_\alpha$ and $U_\alpha$, this is extremely inefficient, and infeasible even for small $n$. 
The problem of efficiently determining $\mathcal I_\alpha$ then reduces to two subproblems. First, minimize the number of permutation tests required to find $L_\alpha$ and $U_\alpha$. Second, minimize the computational effort required to perform a permutation test. 

Regarding the first subproblem, Rigdon and Hudgens showed that $\mathcal I_\alpha$ can be constructed using at most $O(n^4)$ permutation tests \cite{rigdon2015randomization}. Later, Li and Ding showed that the same intervals could be constructed in $O(n^2)$ permutation tests in the balanced case, where an equal number of subjects are assigned to treatment and control (that is, $n = 2m$) \cite{li2016exact}. They also provided numerical evidence that their algorithm reproduces the Rigdon--Hudgens intervals in unbalanced trials  (but did not give a proof of this statement). Specifically, they showed that the interval their algorithm returns always contains the corresponding Rigdon--Hudgens interval $\mathcal I_\alpha$, but did not exclude the possibility that it is strictly larger for some unbalanced designs. 

We now consider the second subproblem. An \emph{exact permutation test} enumerates all $\binom{n}{m}$ possible assignments of subjects, which grows exponentially in $n$ and quickly becomes computationally infeasible. Rigdon and Hudgens therefore recommend performing approximate permutation tests, where the permutation $p$-value is approximated by Monte Carlo simulation with a fixed number of samples $K$ from the distribution of $\bs {\tilde Z}$ in \eqref{intropvalue}. However, such intervals may fail to cover at the nominal rate due to Monte Carlo error. Further, previous literature has not provided any formal analysis regarding the question of how large $K$ should be taken so that the resulting intervals $\mathcal I_\alpha$ cover the unknown parameter $\tau(\bs y)$ with high probability. Such an analysis is not entirely straightforward, since the testing in the Li--Ding algorithm described previously is done in a sequential manner, on a particular subset of the potential outcome tables, and the compatibility of other tables with the observed data inferred from a monotonicity property of the permutation $p$-value.

Our contributions address both of these subproblems.
First, we consider balanced experiments. We analyze this special case in detail because it is common in practice and minimax optimal in terms of estimation efficiency.\footnote{See also \cite[Section 2]{kallus2018optimal} for a justification of balanced randomization via its blinding properties. In general, \cite{kallus2018optimal} shows that balanced complete randomization enjoys a minimax property even when potentially prognostic covariates are available to the designer of the experiment.} 
We show that the confidence intervals $\mathcal I_\alpha$ from \cite{rigdon2015randomization} can be constructed in $4 n \log_2 n$ permutation tests in the balanced case, improving on the $O(n^2)$ tests required by \cite{li2016exact}.\footnote{In this and the next paragraph, we assume that $n \ge 15$ in order to simplify the exposition.} 
The  proof is based on a new monotonicity property for the permutation $p$-values in the balanced case. This permits us to greatly reduce the effective search space when finding $L_\alpha$ and $U_\alpha$.

Further, when the permutation tests are approximated by Monte Carlo simulation with $K$ samples, we show that the resulting intervals cover the unknown parameter $\tau(\bs y)$ with probability at least $1 - \alpha$ if
\be
K\ge \frac{1}{2 \eps^2} \log\left(\frac{4 n \log_2(n) }{\eps} \right),
\ee
where $\eps < \alpha$ is an approximation parameter defined in \Cref{s:montecarlo}, and chosen by the user. Smaller values of $\eps$ correspond to confidence intervals that are smaller on average. 
Observe that a single Monte Carlo sample of $ T(\bs {\tilde Y}, \bs {\tilde Z})$ can be evaluated in $O(n)$ arithmetical operations, by using the Fisher--Yates shuffle algorithm to generate the randomization  $\bs {\tilde Z}$ into equal groups. Then the overall complexity of algorithm producing Monte Carlo intervals for balanced experiments is 
\begin{equation}
O\Bigg(\frac{n^2 \log n }{\eps^2} \log\left(\frac{ n \log (n) }{\eps} \right) \Bigg).
\end{equation}
We further show that by increasing $K$ (and decreasing $\eps$), these intervals will approximate the exact intervals proposed by  \cite{rigdon2015randomization} arbitrarily well, and quantify how $K$ must scale in order to guarantee any desired coverage rate and approximation accuracy as $n$ increases.\footnote{
In our analysis, we treat each fundamental arithmetic operation (addition, subtraction, multiplication, and division) as $O(1)$. This is consistent with the typical convention in computer science; see, for example, \cite[p.\ 26]{cormen2022introduction}.
}

For a general unbalanced trial, certain favorable features of the balanced case are absent, including the monotonicity property noted above. However, we are still able to show that the construction of Rigdon and Hudgens can be improved to require only (the equivalent of) $O(n^2)$ Monte Carlo permutation tests. Unlike the $O(n^2)$ procedure in \cite{li2016exact}, our procedure \textit{provably} returns the intervals in \cite{rigdon2015randomization}, up to Monte Carlo error (which may be made arbitrarily small), and extends to other causal estimands. Our proof is based on a time--memory trade-off, where we reuse permutation test results to determine $p$-values for ``nearby'' potential outcome configurations instead of computing them from scratch. 
In this case, the Monte Carlo intervals cover at the nominal rate regardless of the sample size $K$.


Our algorithmic contributions permit permutation inference to be applied to experiments far larger than those accessible by previous methods. 
Additionally, we extend our analysis to experiments with missing data, a common statistical problem that has not been addressed previously in this context. A highlight of our analysis is that we make no assumption on the missingness mechanism (as opposed to common assumptions such as ``missing at random''). We also provide a generalization of our method for unbalanced trials that permits inference on other causal estimands, including the relative risk ratio and odds ratio.

\subsection{Outline}

We begin in \Cref{s:preliminaries} by introducing our notation and reviewing previous work in detail. In \Cref{s:basic}, we present basic properties of the confidence intervals constructed in \cite{rigdon2015randomization}. In \Cref{s:efficient}, we show how to construct these intervals in $O( n \log n)$ permutation tests in balanced experiments. In \Cref{s:montecarlo}, we continue our analysis of balanced experiments and construct exact intervals using Monte Carlo permutation tests. In \Cref{s:general}, we drop the balanced allocation assumption and demonstrate a confidence interval construction for general experiments that uses $O(n^2)$ Monte Carlo permutation tests. 
We consider missing data in  \Cref{s:missing}, and present simulation evidence in \Cref{s:simulations}. \Cref{s:open} discusses some remaining open problems. 
Appendices A through E contain the proofs for Sections 3 through 7, respectively. 
\subsection{Code}
Code implementing our proposed algorithms and replicating the tables provided in \Cref{s:simulations} is available at \url{https://github.com/hchang1508/BinaryOutcomePermutationTest_replication}.

\section{Preliminaries}\label{s:preliminaries}

\subsection{Notation}
Let $\mathbb{Z}_{>0}$ denote the positive integers. 
We consider a group of $n \in \mathbb{Z}_{>0}$ subjects who undergo an experiment with a binary outcome. 
We let $m$ denote the number of subjects assigned to the treatment group. Then $n-m$ subjects are assigned to control. We let $\unn{a}{b}$ denote the set $\{ k \in \mathbb{Z} : a \le k \le b \}$.

A \emph{potential outcome table} is any vector 
\be\label{e:wdef}
\bs w \in \big\{ (0,0), (0,1), (1,0), (1,1) \big\}^n.
\ee
Letting $\boldsymbol{w}_i = ( w_i(1), w_i(0) )$ for $i \in \unn{1}{n}$, the coordinate $w_i(1)$ denotes the outcome for the $i$-th subject under treatment, and $w_i(0)$ denotes the outcome under control. We use $\boldsymbol{y}$ to denote the potential outcome table for the subjects in the experiment, which is a fixed (but unknown) ground truth. 
For every table $\bs w$, we let the corresponding count vector $\bs v  = \bs v (\bs w) \in \mathbb{Z}_{\ge 0}^{\{0,1\}^2}$ be defined by
\be\label{vdef}
\bs v = (v_{11}, v_{10}, v_{01}, v_{00}) , \qquad  v_{ab} = \sum_{j=1}^n \one\{ \bs w_j = (a,b) \}.
\ee

We now introduce notation for the randomization of the subjects. Consider the random vector 
\be
\boldsymbol{Z} = (Z_1, Z_1, \dots, Z_n) \in \{0,1\}^n
\ee
whose  distribution is uniform over the set
\be\label{zn}
\mathcal Z(n) = \left\{ \bs z \in \{0,1\}^n  : \sum_{i=1}^n z_i = m \right\} .
\ee
Here, the indices $i$ such that $Z_i = 1$ represent the subjects assigned to treatment. The vector of observed experimental outcomes $\bs Y = (Y_j)_{j=1}^n$ is then given by
\be\label{bigY}
Y_j = Z_j y_j(1) + (1 -  Z_j) y_j(0).
\ee
We say that a table $\bs w$ is \emph{possible} given the observed data $\bs Y$ if for every $j \in \unn{1}{n}$, we have $\bs w_j(1) = Y_j$ if $Z_j=1$, and $\bs w_j(0) = Y_j$ if $Z_j = 0$. We say that a count vector $\bs v$ is possible if it arises from some possible table $\bs w$.


The sample average treatment effect for a potential outcome table is defined by
\be
\tau(\bs{w}) = \frac{1}{n} \sum_{j=1}^n \big( w_j(1) - w_j(0) \big).
\ee
The Neyman estimator of $\tau(\bs y)$ is given by 
\be
T(\bs Y, \bs Z) = \sum_{j=1}^n  \frac{Z_j Y_j}{m}  - \sum_{j=1}^n \frac{(1-Z_j) Y_j}{n-m}.
\ee
This estimator depends on $(\bs Y, \bs Z)$ only through the observed count vector $\bs n \in \mathbb{Z}_{\ge 0}^{\{0,1\}^2}$ defined by 
\be\label{ndef}
\bs n = (n_{11}, n_{10}, n_{01}, n_{00}), \qquad  n_{zy} = \sum_{j=1}^n \one\{ Z_j = z, Y_j = y  \},
\ee
for $z\in\{0,1\}$ and $y \in \{0,1\}$. For example, $n_{10}$ is the number of subjects in the treatment group with observed outcome $0$. Then \[T(\bs Y, \bs Z)=\frac{n_{11}}{m}-\frac{n_{01}}{n-m}.\]
We overload the notation slightly and write $T(\bs n)$ for the value of the Neyman estimator $T(\bs Y, \bs Z)$, if $\bs n$ is the observed count vector for $(\bs Y, \bs Z)$. 
Below, there will be several other functions of $(\bs Y, \bs Z)$ that similarly depend on $(\bs Y, \bs Z)$ only through its observed count vector $\bs n$, and we will overload their notation in a parallel way. These include $\mathcal I_\alpha$ (see \eqref{e:shorthand}), $\mathcal I_{\alpha, \eps, K}$ (see \Cref{s:montecarlo}), and $\mathcal I_{\alpha, K}$ (see the first paragraph of \Cref{s:IaKanalysis}).

We write $\log$ for the natural logarithm. When we use the base 2 logarithm, it is always denoted by $\log_2$. Additionally, we often abbreviate probability mass function as pmf. Further notation, used only in the proofs, is introduced in \Cref{s:basicproofs}.


\subsection{Previous Work}

\subsubsection{Rigdon and Hudgens}

We begin by recalling the confidence interval procedure proposed in \cite{rigdon2015randomization}. 
As $\bs w$ ranges over all tables, $\tau(\bs w)$ takes every value in the set 
\be
\mathcal S(n) = \left\{ - \frac{n}{n}, -\frac{n-1}{n}, \dots, \frac{0}{n} , \dots , \frac{n-1}{n}, \frac{n}{n}  \right\},
\ee
which has $2n+1$ elements. After the experiment is completed, the set of $\tau(\bs w)$ for $\bs w$ that are possible given the observed data is further restricted to 
\be\label{mathcalC}
\mathcal C (\bs Y, \bs Z) 
= \left\{ \frac{1}{n}\left(\sum_{j=1}^n (2Z_j -1 )Y_j - m \right), \dots, 
\frac{1}{n}\left(\sum_{j=1}^n (2Z_j -1 )Y_j - m  + n \right)
\right\},
\ee
which has $n+1$ elements. We observe that $\mathcal C$ depends on $(\bs Y, \bs Z)$ only through the associated observed count vector $\bs n$.

We now consider some $\alpha \in (0,1)$ and a realization of $(\bs Y, \bs Z)$. We will construct a confidence interval for $\tau\left(\bs y\right)$ at level $1-\alpha$ (that is, one that contains the true parameter with probability at least $1-\alpha$). We note that $\tau\left(\bs y\right)$ is a function solely of the count vector $\bs v$ associated with the potential outcome table $\bs y$.
We first describe a permutation test for the sharp null hypothesis that $\bs y = \bs w$ for some given $\bs w$, then explain how to form a confidence interval from a series of these tests.

\begin{definition}\label{d:permtest}
Let $\bs{\widetilde Z}$ be a random vector independent from $\bs Z$ with the same distribution, and let $\bs{\tilde Y}$ be the vector with entries 
\be
 {\tilde Y}_j = \tilde Z_j w_j(1) + (1 - \tilde Z_j) w_j(0).
\ee
We say that $\bs w$  is compatible with observed data $(\bs Y, \bs Z)$ at level $1-\alpha$ if\footnote{
The inequality $\ge \alpha$ can be replaced with the inequality $>\alpha$ in \eqref{pvalue}, which creates a stricter threshold for acceptance. This change leads to potentially shorter confidence intervals  that still cover at the nominal rate, which can be seen by inspecting the proof of \Cref{p:coverage}. Further, all of our results are still valid under the stricter definition. The definition \eqref{pvalue} was used in the previous works \cite{rigdon2015randomization,li2016exact}, and we retain it 
for consistency.}
\be\label{pvalue}
p(\bs w, \bs Y , \bs Z) = \P\left( \big| T(\bs{\tilde Y}, \bs{ \tilde Z} )  - \tau(\bs w) \big| \ge \big| T(\bs Y, \bs Z )  - \tau(\bs w) \big| \right) \ge \alpha.
\ee
We say that $\bs w$ is compatible with an observed count vector $\bs n$ if $\eqref{pvalue}$ holds with $T(\bs Y,\bs Z)$ replaced by $T(\bs n)$.
\end{definition}
Let $\bs  v$ denote the count vector associated with $\bs w$, as defined in \eqref{vdef}, and let $\bs n$ be as in \eqref{ndef}. Then it is straightforward to see that $p(\bs w, \bs Y , \bs Z)$ 
depends on $(\bs w, \bs Y , \bs Z)$ only through the count vectors $\bs n$ and $\bs v$. This motivates the shorthand $p (\bs v, \bs n) = p(\bs w, \bs Y , \bs Z)$. 
\begin{alg}\label{a:permutation}
A \emph{permutation test} takes as inputs $\alpha \in (0,1)$, a  count vector $\bs v$, and a vector $\bs n \in \mathbb{Z}_{\ge 0}^{\{0,1\}^2}$. It returns a binary decision, accept or reject, as follows. The algorithm computes $p(\bs v, \bs n)$ by direct enumeration of all $\binom{n}{m}$ elements of the set $\mathcal Z(n)$ (defined in \eqref{zn}), over which $\tilde {\bs Z}$ is uniformly distributed. If $p(\bs v, \bs n) \ge \alpha$, it accepts. Otherwise, it rejects. 
\end{alg}

We now present the confidence interval procedure from \cite{rigdon2015randomization}.  Given an observed count vector $\bs n$, set
\be
\mathcal T(\bs n) = \big\{
(i,j,k,l) : i \in \unn{0}{n_{11}}, j\in \unn{0}{n_{10}}, k \in \unn{0}{n_{01}}, l \in \unn{0}{n_{00}}
\big\}.
\ee
 We define vectors $\bs v(i,j,k,l)$ in the following way.
Given $(i,j,k,l) \in \mathcal T(\bs n)$, set
\be
\bs v(i,j,k,l) = ( i + k, n_{11} - i + l,
n_{01} -k +j, n_{10} +n_{00} - j - l ).
\ee
To motivate this definition, we think of starting with the observed count vector $\bs n$, and imputing various combinations of the unknown potential outcomes. In the $n_{11}$ subjects with observed outcome 1 in the treatment group, we impute $1$ for the unknown outcomes in exactly $i$ of them, to get $i$ potential outcome pairs $(1,1)$. Similarly, $j$, $k$, and $l$ represent the number of $1$'s imputed in the other three combinations of group and outcome. It is straightforward to see that the collection of $\bs v(i,j,k,l)$ such that $(i,j,k,l) \in \mathcal T(\bs n)$ is exactly the set of count vectors $\bs v$ possible given the observed count vector $\bs n$. (However, the map $(i,j,k,l) \mapsto  \bs v(i,j,k,l)$ is not necessarily injective.)

\begin{alg}\label{a:rh}
This algorithm takes as input $\alpha \in (0,1)$ and a vector $\bs n \in \mathbb{Z}_{\ge 0}^{\{0,1\}^2}$. It returns an interval $\mathcal I_\alpha(\bs n) = \big[ L_\alpha(\bs n), U_\alpha(\bs n) \big]$. 

By directly enumerating of all elements of $\mathcal T$, and applying \Cref{a:permutation} to each one, compute the set 
\be
\mathcal K (\bs n)
= \Big\{ \tau\big(\bs v(i,j,k,l)\big) : (i,j,k,l)\in \mathcal T (\bs n)
\text{ and \Cref{a:permutation} accepts  $\bs v(i,j,k,l)$}
\Big\}.
\ee
Return $U_\alpha(\bs n) = \max( \mathcal K)$ and $L_\alpha = \min(\mathcal K)$. 
\end{alg}

As noted previously, any observed data $(\bs Y, \bs Z)$ gives rise to an observed count vector $\bs n$. Then using \Cref{a:rh}, we define 
\be\label{e:shorthand}
\mathcal I_\alpha (\bs Y, \bs Z) = \big[L_\alpha(\bs Y, \bs Z), U_\alpha(\bs Y, \bs Z)\big]
\ee
by setting $\mathcal I_\alpha (\bs Y, \bs Z) = \mathcal I_\alpha ( \bs n)$.
Rigdon and Hudgens observe that since $\mathcal T(\bs n)$ has size $O(n^4)$, only $O(n^4)$ permutation tests are needed to construct 
the set $\mathcal K(\bs n)$ in \Cref{a:rh}. Further, this algorithm does not require that the treatment and control groups have an equal number of subjects, and may be applied to arbitrary unbalanced trials.

\subsubsection{Li and Ding}
As noted in the introduction, Li and Ding have proposed an algorithm that returns $\mathcal I_\alpha(\bs n)$ in  $O(n^2)$ permutation tests for balanced trials \cite{li2016exact}. We do not give the details here, since they are not relevant to our work. However, we will use the following lemmas from \cite{li2016exact}. The first provides a  necessary and sufficient condition for checking whether a potential outcome table is possible given the observed data. It does not require that the design be balanced.

\begin{lem}[{\cite[Theorem 1]{li2016exact}}]\label{l:ispossible}
A potential outcome table $\bs w$ with count vector $\bs v$ is possible given an observed count vector $\bs n$ if and only if
\begin{align}
&\max( 0,  n_{11} -  v_{10},
 v_{11} - n_{01},
 v_{11} + v_{01} - n_{10} - n_{01})\\
&\le
\min(
 v_{11}, n_{11},  v_{11} + v_{01} - n_{01},
 n - v_{10} - n_{01} - n_{10} \notag
).
\end{align}
\end{lem}

The second lemma shows that in the balanced case, the confidence set that we would obtain if we tested \emph{every} possible value of $\tau_0$ is indeed an interval. 
\begin{lem}[{\cite[Theorem A.4]{li2016exact}}]\label{l:isinterval}
Suppose $n=2m$. Fix $\alpha \in (0,1)$ and observed data $\bs Y$ and $\bs Z$. For every $\tau_0 \in \mathcal I_\alpha (\bs Y, \bs Z) \cap \mathcal C$, there exists a possible potential outcome table $\bs w$ such that $\tau(\bs w) = \tau_0$ and $p(\bs w, \bs Y, \bs Z) \ge \alpha$.
\end{lem}






\section{Basic Properties}\label{s:basic}

We now note three basic properties of the confidence intervals $\mathcal I_\alpha$. While the first two have been mentioned informally in previous works, they have not been proved, and we feel there is value in giving precise statements and justifications. The third is new. The proofs appear in \Cref{s:basicproofs}. 

Our first lemma implies that the interval $\mathcal I_\alpha(\bs n)$ always contains the estimate $T(\bs n)$ of $\tau(\bs y)$ for balanced trials.\footnote{This claim is not always true for unbalanced trials. For example, consider an experiment with $9$ units, where $7$ are assigned to treatment and the true potential outcome table  is $v_{11}=0$, $v_{10}=1$, $v_{01}=0$, and $v_{00}=8$. Suppose we observe $n_{11}=1$, $n_{10}=6$, $n_{01}=0$, and $n_{00}=2$, and therefore $T(\bs n)=1/7$. Setting $\alpha=17/18$, it can be shown by direct combinatorial analysis (or simulation) that only $-1/9$, $0$, and $1/9$ are accepted. Then the realized confidence interval is $[-1/9, 1/9]$, excluding the observed estimate of $1/7$.}
\begin{lem}\label{l:containsobserved}
Suppose $n = 2m$. For every $\bs n \in \mathbb{Z}_{\ge 0}^{\{0,1\}^2}$, we have 
\be L_\alpha(\bs n) \le T(\bs n) \le U_\alpha(\bs n).\ee 
\end{lem}
The next proposition states that the intervals $\mathcal I_\alpha$ are exact.
It holds in the general unbalanced setting. Given observed data $(\bs Y, \bs Z)$, we recall that  the interval $\mathcal I_{\alpha}(\bs Y, \bs Z) $ was defined in \eqref{e:shorthand}.

\begin{prop}\label{p:coverage}
Fix $n,m \in \mathbb Z_{>0}$ with $m < n$, $\alpha \in (0,1)$, and true potential outcomes $\bs y$. Then 
\be
\P\big ( \tau(\bs y) \in \mathcal I_\alpha(\bs Y , \bs Z) \big ) \ge 1 - \alpha,
\ee
where the probability is with respect to the variable $\bs Z$.
\end{prop} 

Finally, the following proposition states that the intervals $\mathcal I_\alpha$ converge at a $n^{-1/2}$ rate when $m$ is proportional to $n$. This rate is characteristic of many confidence interval procedures based on central limit theorem asymptotics. We therefore see that the large $n$ scaling behavior of the length of the interval $\mathcal I_\alpha$ is the same as that of the intervals produced by asymptotic methods when $m/n$ is bounded away from from $0$ and $1$.\footnote{
In \Cref{rate}, our focus is on establishing the correct scaling rate, not obtaining the optimal constant prefactor. This estimate is quite conservative compared to the interval lengths we find empirically through simulation. 
A sharper bound for the balanced design, derived using the Chernoff bound, is available in an earlier version of the paper \cite{aronow2023fast}.
}  

\begin{prop}\label{rate}
Fix $n,m \in \mathbb Z_{>0}$ with $m < n$, $\alpha \in (0,1)$, and true potential outcomes $\bs y$. 
For every $\bs n \in \mathbb{Z}_{\ge 0}^{\{0,1\}^2}$, we have
\be
\big| I_\alpha(\bs n) \big| \le  \sqrt{\frac{4}{\alpha} \cdot \frac{n }{n-1} \cdot \frac{n}{ m (n-m)} } .
\ee
\end{prop}




\section{Balanced Experiments}\label{s:efficient}
In this section, we construct confidence intervals for balanced experiments using both exact permutation tests (\Cref{a:permutation}) and approximate Monte Carlo permutation tests.

\subsection{Exact Intervals}
We now present our result on the efficient computation of $\mathcal I_\alpha$ in the balanced case. Its proof is given in \Cref{s:mainproof}.
\begin{thm}\label{t:main}
Suppose $n\ge 15$ and $n=2m$.
For every $\alpha \in (0,1)$ and observed count vector $\bs n$, the interval $\mathcal I_\alpha(\bs n)$ can be constructed using at most $4 n \log_2 n $ permutation tests.
\end{thm}

To  prove \Cref{t:main}, we exhibit an algorithm that constructs $\mathcal I_\alpha(\bs n)$ in the required number of permutation tests.\footnote{Again, the estimate in \Cref{t:main} is conservative relative to the empirical performance of \Cref{a:efficient} below. To streamline the arguments, we did not pursue the sharpest possible bounds.} 
We begin with a heuristic argument. We first observe that the problem of finding the upper bound $U_\alpha(\bs n)$ reduces to finding a method to determine whether a given value $\tau_0 \in \mathcal C$ is incompatible with the observed data $\bs n $, in the sense that $p(\bs v, \bs n) < \alpha$ for all  $\bs v$ possible given $\bs n$ such that $\tau(\bs v) = \tau_0$.
If this can be done in $O(n)$ permutation tests, then we can perform a binary search on the set $[T(\bs n), \max ( C(\bs n) ) ]\cap \mathcal S(n)$ to find $U_\alpha$ by testing each value $\tau_0$ considered by the binary search for compatibility with $\bs n$.
Since \Cref{l:isinterval} guarantees that every element of $\mathcal I_\alpha(\bs n)$ is compatible with the observed data, and \Cref{l:containsobserved} guarantees that $\mathcal I_\alpha(\bs n)$ contains $T(\bs n)$, the binary search will return the maximum of 
\be
\Big[T(\bs n), \max \big( C(\bs n) \big) \Big]\cap \mathcal S(n) \cap \mathcal I_\alpha(\bs n),
\ee
which is equal to $U_\alpha(\bs n)$. Since binary search on a set of size $O(n)$ can be completed in $O(\log n )$ time, in total $O( n \log n)$ permutation tests are needed. Analogous reasoning applies for finding the lower bound $L_\alpha$.


Consequently, the proof of \Cref{t:main} focuses on showing that a given $\tau_0 \in \mathcal C$ can be checked for compatibility in $O(n)$ permutation tests. 
Observe that any possible count vector $\bs v = (v_{11}, v_{10}, v_{01}, v_{00})$ such that $\tau(\bs v) = \tau_0$ satisfies the equations 
\be\label{a1}
v_{11} +  v_{10} +  v_{01} +  v_{00} = n,\qquad v_{10} -  v_{01} = n \tau_0.
\ee
The following lemma allows us to efficiently search the space cut out by \eqref{a1} by exploiting a certain monotonicity property of the permutation $p$-values defined in \eqref{pvalue}. It is proved in \Cref{s:monotonic}.

\begin{lem}\label{l:monotonic} Suppose $n=2m$. Fix observed data $\bs n$, and a count vector $
\bs v = (v_{11}, v_{10}, v_{01}, v_{00})$.
Suppose $\min(v_{10}, v_{01}) \ge 1$ and $\max(v_{10}, v_{01}) \ge 2$, and set
\be\label{a3}
\bs v' = (v_{11}+1, v_{10}-1, v_{01}-1, v_{00}+1).
\ee
Then $
p( \bs v', \bs n) \ge p(\bs v, \bs n).$
\end{lem}
The upshot of this lemma is that if $p(\bs v, \bs n) < \alpha$, meaning $\bs v$ is incompatible with the observed data, then  all potential outcome tables on the line segment given by the translations \eqref{a3}, with endpoint $\bs v$, are also incompatible. Heuristically, the transformation in \eqref{a3} causes the distribution of the variable $T(\bs{\tilde Y}, \bs{ \tilde Z} )  - \tau(\bs w)$ in the definition \eqref{pvalue} to become more spread out about its mean (zero), increasing the $p$-value. This heuristic is made precise in the proof of \Cref{l:monotonic}.

Given this context, we now present our main algorithm. The additional check in Step~2(c) exists to handle the case where \Cref{l:monotonic} does not apply. 
\begin{alg}\label{a:efficient}
This algorithm takes as input $\alpha \in (0,1)$ and a vector $\bs n \in \mathbb{Z}_{\ge 0}^{\{0,1\}^2}$. It returns an interval $[ L^{(1)}_\alpha(\bs n),  U^{(1)}_\alpha(\bs n) ]$. We give only the steps for finding $U^{(1)}_\alpha(\bs n)$, since finding $ L^{(1)}_\alpha(\bs n)$ is analogous.

We perform a binary search using a function
$f \colon \unn{n T(\bs n )}{ n \max( \mathcal C (\bs n))} \rightarrow \{0,1\}$ defined below. The precise definition of this binary search is given as \Cref{a:binary} in \Cref{s:eprelim}.
Given an input $x$, the function $f$ returns $0$ if $x/n$ is compatible with the vector $\bs n$, and $1$ otherwise, where the  compatibility of a given $\tau_0 \in \mathcal C(\bs n)$ is determined through the following steps.

\begin{enumerate}
\item Initialize $j\leftarrow 0$.
\item\label{step2} If $j \le n$, perform the following steps.
\begin{enumerate}

\item\label{stepa} 
Construct a count vector $\bs v$ as follows (recall \eqref{vdef}). 
Let $v_{10}$ be a free parameter, and set $v_{11} = j - v_{10}$. Then (solving \eqref{a1}) we set
\be\label{constructv}
v_{01} = v_{10} - n \tau_0,\qquad 
v_{00} = n - j - v_{10} + n \tau_0.
\ee

\item \label{stepb} 
Find the smallest $v_{10}$ that leads to a possible table $\bs v$ given the observed data $\bs n$. This can be done in constant time using  \Cref{l:possibleinequalities}, stated in \Cref{s:eprelim} below.
If no value of $v_{10}$ leads to a possible table, set $j \leftarrow j +1$ and go to \eqref{step2}. 

\item\label{step5} Permutation test the count vector $\bs v$ coming from the choice of $v_{10}$ in the previous step using \Cref{a:permutation} with the given value of $\alpha$.  If $\bs v$ is accepted, return that $\tau_0$ is compatible. 
If $\bs v$ is rejected and $v_{10} \ge 1$ or $v_{01} \ge 1$,  set $j \leftarrow j +1$ and go to \eqref{step2}. 

If $\bs v$ is rejected and $v_{10} = v_{01} = 0$, additionally permutation test the count vector given by $v_{10} = 1$. If it is possible and accepted, terminate the algorithm and declare $\tau_0$ compatible. Otherwise, set $j \leftarrow j +1$ and go to \eqref{step2}.
\end{enumerate} 
\item Return that $\tau_0$ is incompatible.
\end{enumerate}
\end{alg}



\subsection{Monte Carlo Intervals}\label{s:montecarlo}

We now construct confidence intervals using approximate permutation tests, which are far less computationally demanding than exact tests when $n$ is large. 
The following algorithm approximates the true permutation $p$-values by Monte Carlo simulation. It includes an additional $\eps$ term in the rejection threshold to guard against problems that would otherwise arise when the true $p$-value is close to the level $\alpha$. 

Given a count vector $\bs v$, we define $\tilde {\bs w} = \tilde {\bs w}(\bs v)$ as the potential outcome table such that 
\begin{align}
\begin{split}\label{tildew}
\tilde {\bs w}_j &= (1,1) , \quad j=1,\dots, v_{11},\\
\tilde {\bs w}_j &= (1,0) , \quad j=v_{11} + 1 ,\dots, v_{11} + v_{10},\\
\tilde {\bs w}_j &= (0,1) , \quad j=v_{11} + v_{10} +  1 ,\dots, v_{11} + v_{10} + v_{01},\\
\tilde {\bs w}_j &= (0,0) , \quad j= v_{11} + v_{10} + v_{01} + 1, \dots, n.
\end{split}
\end{align}
The map $\tilde {\bs w}$ is an arbitrary choice of function taking a count vector $\bs v$ to a table with that count vector. We fix this choice for concreteness, but any other would work just as well. 

\begin{alg}\label{a:approximatetest}
Take as input $\alpha \in (0,1)$, $\eps \in [0, \alpha)$, $K \in \mathbb{Z}_{> 0}$,  a count vector $\bs v \in \mathbb{Z}_{\ge 0}^{\{0,1\}^2}$, and a vector $\bs n \in \mathbb{Z}_{\ge 0}^{\{0,1\}^2}$. The algorithm will return a binary decision, either accept or reject.

Let $\bs Z^{(1)}, \dots , \bs Z ^{(K)}$ be a sequence of independent, identically distributed random vectors, with common distribution equal to that of $\bs Z$. 
Define random variables $V_1, \dots, V_K$ by
\be
V_i = \one \Big( \big| T (\bs{n}^{(i)} )  - \tau(\bs {\tilde w}) \big| \ge \big| T(\bs n )  - \tau(\bs {\tilde w}) \big| \Big), \qquad i \in \unn{1}{K},
\ee 
where $\bs n^{(i)}$ is the outcome count vector corresponding to $(\bs Y^{(i)}, \bs Z^{(i)} )$, with  $\bs Y^{(i)}$ defined by
\be
 Y^{(i)}_j =  Z^{(i)}_j \tilde w_j(1) + (1 -  Z^{(i)}_j) \tilde w_j(0)
\ee
for all $j \in \unn{1}{n}$. 
Set
\be\label{approxpval}
S = \frac{1}{K} \sum_{i=1}^K V_i.
\ee
Simulate a realization of $S$ by simulating realizations of the $K$ variables $\bs Z^{(1)}, \dots , \bs Z ^{(K)}$, and constructing the $V_i$ and $S$ accordingly. 
Return the decision accept if $S+\eps \ge \alpha$, and return the decision reject otherwise.
\end{alg}

\begin{alg}\label{a:approximate}
This algorithm takes as input $\alpha \in (0,1)$, $\eps \in [0, \alpha)$, and a vector $\bs n \in \mathbb{Z}_{\ge 0}^{\{0,1\}^2}$. It returns an interval $\mathcal I_{\alpha, \eps, K}(\bs n)  = \big[ L_{\alpha, \eps, K}(\bs n),  U_{\alpha, \eps, K} (\bs n) \big]$. The algorithm is identical to \Cref{s:efficient}, except every use of an exact permutation test (\Cref{a:permutation}) is replaced with an approximate permutation test with the given parameters (\Cref{a:approximatetest}). A new set of independent vectors $\bs Z^{(1)},\dots \bs Z^{(k)}$ is generated for each approximate permutation test (independently of the permutation vectors used for any previous tests).
\end{alg}

Given observed data $(\bs Y, \bs Z)$ we define the interval $\mathcal I_{\alpha, \eps, K}(\bs Y, \bs Z) $ to be the interval $\mathcal I_{\alpha, \eps, K}(\bs n)$ returned by \Cref{a:approximate} for the observed count vector $\bs n$ corresponding to $(\bs Y, \bs Z)$.

The first part of the next proposition shows that the intervals generated by \Cref{a:approximate} are guaranteed to cover $\tau(\bs y)$ with probability at least $1-\alpha$, if the parameters of the algorithm are chosen correctly. The second part shows that these intervals can be made to approximate the deterministic intervals $\mathcal I_\alpha$ arbitrarily well. The proof is deferred to \Cref{s:montecarloproof}. 

\begin{prop}\label{p:montecarlo}
Suppose $n=2m$. Fix a potential outcome table $\bs y$, $\alpha \in (0,1)$, and $\eps \in (0,1)$.  Suppose $n\ge 15$. 
\begin{enumerate}
\item 
Fix $\eps \in (0, \alpha)$.
If 
\be \label{K}
K\ge \frac{1}{2\eps^2} \log\left(\frac{4 n \log_2(n) }{\eps} \right),\ee  then 
\be
\P\big ( \tau(\bs y) \in \mathcal I_{\alpha-\eps, \eps, K}(\bs Y , \bs Z) \big ) \ge 1 - \alpha.
\ee
\item Fix $\eps \in (0, \alpha/3)$. We have 
\be
\P\big(\mathcal I_{\alpha-\eps, \eps, K}(\bs Y , \bs Z)  \subset \mathcal I_{\alpha -3 \eps} ) \ge 1 -   4  n \log_2(n) \exp(-2K\eps^2) . 
\ee 
\end{enumerate}
The probability statements are with respect to joint distribution of  $\bs Z$ and the random Monte Carlo draws $\bs Z^{(i)}$ used in each application of \Cref{a:approximatetest}.
\end{prop}

\Cref{p:montecarlo} clarifies two important points. First, if $n$ is held fixed, guaranteeing coverage at level $1-\alpha$ for the $\eps$-approximate interval  $\mathcal I_{\alpha - \eps, \eps, K}$ requires a  number of Monte Carlo samples $K$ that scales asymptotically as $\eps^{-2} \log ( \eps^{-1})$ as $\eps$ tends to zero.  Second, if $\eps$ is fixed, the  number of samples $K$ should increase asymptotically as $\log( 4 n \log_2(n))$  as $n$ grows to guarantee that the Monte Carlo intervals are contained in the exact interval $\mathcal I_{\alpha -3 \eps}$.

We note that \Cref{a:approximate} is trivially parallelizable, for example by dividing the computation of the $V_i$ in \eqref{a:approximatetest} across distinct processors. 
\begin{remark}
For simplicity, we have stated \Cref{p:montecarlo} using a fixed number of $K$ samples for each approximate permutation test. This number is conservative, since it is chosen to ensure that the approximate tests return the correct decision for $p$-values near the threshold $\alpha$ (with high probability). It may be more efficient to use an adaptive strategy with a variable number of samples $K$, and to stop sampling variables $V_i$ and reject $\bs w$ when $\E[V_1]+\eps \ge \alpha$ can be excluded with sufficiently high confidence. We conjecture that the validity of such an  algorithm could be justified by replacing the concentration inequalities used in the proof of \Cref{p:montecarlo} with more general ones that account for the correlations induced by the early stopping. Such inequalities are well known in the bandit algorithm literature; see for example \cite[Exercise 7.1]{lattimore2020bandit}. However, for brevity, we do not analyze this method here, leaving it to future work.

A practically useful, but less formal, version of this idea is to first run \Cref{a:approximate} with a moderate value of $K$ (say $K=2000$), $\eps = 0$, and the desired value of $\alpha$, and view the result $\mathcal I_{\alpha, 0, K}$ as an approximation to the true interval $\mathcal I_{\alpha}$. Then by testing values of $\tau_0$ near the returned endpoints $L_{\alpha, 0, K}$ and $U_{\alpha, 0, K}$, using Steps 1 through 3 of \Cref{a:approximate} with a larger number of Monte Carlo samples, this approximation may be refined to arbitrarily high accuracy until the analyst is satisfied that it matches the true $\mathcal I_{\alpha}$.\footnote{When the true $p$-value is exactly equal to $\alpha$, successive Monte Carlo approximations will never be dispositive, as they will fluctuate above and below $\alpha$ indefinitely as the sample size grows. This problem rarely arises in practice, and can be resolved by accepting the corresponding table when there is ambiguity, to ensure the confidence interval remains valid.} This method is also applicable to the more general Monte Carlo algorithm in the next section.
\end{remark}

\section{General Experiments}\label{s:general}
The computation of the intervals $\mathcal I_\alpha$ in the general case presents two major challenges compared to the balanced one. First, \Cref{l:isinterval} may not hold, which prevents us from using a binary search as in \Cref{a:efficient}. Second, the conclusion of \Cref{l:monotonic} may not hold. In this section, we present a new algorithm for approximate confidence intervals that accommodates general causal estimands, does not require a balanced design, and has time complexity equivalent to $O(n^2)$ Monte Carlo permutation tests. We then specialize it to the estimation of the sample average treatment effect, where a slight gain in efficiency is possible.

\subsection{Algorithm}

We begin by making a more general definition of an approximate permutation test, which allows for any \emph{estimand} $\theta\colon \mathbb{Z}_{\ge 0}^{\{0,1\}^2} \rightarrow \mathbb{R}$ and \emph{test statistic} $R \colon \mathbb{Z}_{\ge 0}^{\{0,1\}^2} \times  \mathbb{Z}_{\ge 0}^{\{0,1\}^2} \rightarrow \mathbb R$.  Estimands $\theta$ will take potential outcome count vectors $\bs v$ as arguments, and test statistics $R$ will take pairs $(\bs n, \bs v)$ of observed count vectors $\bs n$ and count vectors $\bs v$ as arguments. 

\begin{alg}\label{a:approximatetest2}
Take as input $\alpha \in (0,1)$, 
$K \in \mathbb{Z}_{> 0}$, an estimand $\theta$, a test statistic $R$, a count vector $\bs v \in \mathbb{Z}_{\ge 0}^{\{0,1\}^2}$, and a vector $\bs n \in \mathbb{Z}_{\ge 0}^{\{0,1\}^2}$. The algorithm will return a binary decision, either accept or reject, the value $\theta(\bs v)$, and  $K$ vectors lying in $\mathcal Z (n)$. 

Let $\bs Z^{(1)}, \dots , \bs Z ^{(K)}$ be a sequence of independent, identically distributed random vectors, with common distribution equal to that of $\bs Z$. 
Define random variables $V_1, \dots, V_K$ by
\be\label{videf}
V_i = \one \Big( \big| R (\bs{n}^{(i)}, \bs v  )  \big| \ge \big| R(\bs n ,\bs v)   \big| \Big), \qquad i \in \unn{1}{K},
\ee 
where $\bs n^{(i)}$ is the outcome count vector corresponding to $(\bs Y^{(i)}, \bs Z^{(i)} )$, with  $\bs Y^{(i)}$ defined by
\be
 Y^{(i)}_j =  Z^{(i)}_j \tilde w_j(1) + (1 -  Z^{(i)}_j) \tilde w_j(0)
\ee
for all $j \in \unn{1}{n}$, and we recall the definition of $\tilde {\bs w}$ from \eqref{tildew}. 
Set
\be\label{approxpval2}
S = \frac{1}{K+1} \left( 1 + \sum_{i=1}^K V_i \right).
\ee
Simulate a realization of $S$ by simulating realizations of the $K$ variables $\bs Z^{(1)}, \dots , \bs Z ^{(K)}$, and constructing the $V_i$ and $S$ accordingly. 
Return the decision accept if $S > \alpha$, and return the decision reject otherwise. Also return $\theta(\bs v)$ and the realized values of $\bs Z^{(1)}, \dots , \bs Z ^{(K)}$.
\end{alg}

The definition \eqref{approxpval2} can be thought of as counting the original observation $R(\bs n, \bs v)$ as a fictitious $(K+1)$-th randomization, for which the indicator function in \eqref{videf} is always equal to one. 

\begin{remark}\label{r:pexact}
We make a small change in \eqref{approxpval2} relative to the previous definition to take advantage of the well-known fact that that it makes $S$ a valid $p$-value for testing the null hypothesis that the count vector of $  \bs y$  is $\bs v$ (where we recall that $\bs y$ denotes the true potential outcome table). More precisely, under this null, 
\be\label{e:validp}
\P(S \le \alpha ) \le \alpha
\ee 
for all $\alpha \in [0,1]$, 
where the probability is taken with respect to the joint distribution of $\bs Z$ and 
$\bs Z^{(1)}, \dots , \bs Z ^{(K)}$. 
This follows from the fact that when the null is true, the $K+1$ random variables 
\be
R(\bs n , \bs v) , 
R (\bs{n}^{(1)}, \bs v ) ,
\dots,
R (\bs{n}^{(K)}, \bs v  ) 
\ee 
are identically distributed  \cite[Theorem 1]{harrison2012conservative}. 
\end{remark}

\Cref{a:approximatetest2} is flexible enough to accommodate many test statistics and estimands used in practice.

\begin{itemize}
\item To target the sample average treatment effect, we have already seen that we may use the estimand
\be
\theta (\bs v)=\tau (\bs v) = \frac{v_{11} - v_{01}}{n} = \frac{v_{11} - v_{01}} {v_{11} + v_{10} + v_{01} + v_{00}}.
\ee
and the test statistic $T(\bs n) - \theta(\bs v)$, where 
\be
T(\bs n) = \frac{ n_{11} }{n_{11} + n_{10} } - \frac{n_{01}}{n_{01} + n_{00}}.
\ee 
\item For inference on $\tau$, we could also use the studentized test statistic 
\be
R_{\mathrm{std}}( \bs n, \bs v )  = \frac{ T(\bs n) - \tau(\bs v)} {\sigma(\bs n) },
\ee
where
\begin{align}
\sigma^2(\bs n) &= \frac{Q_1}
{(n_{11} + n_{10})^2} 
+ 
\frac{Q_0}{(n_{01} + n_{00})^2},\\
Q_1 &= n_{11}  
\left( 
\frac{n_{11}}{n_{11} + n_{10}}  -1  
\right)^2  + n_{10} \left( 
\frac{n_{11}}{n_{11} + n_{10}} 
\right)^2,\\
Q_0 &= n_{01}  
\left( 
\frac{n_{01}}{n_{01} + n_{00}}  -1  
\right)^2  + n_{00} \left( 
\frac{n_{01}}{n_{01} + n_{00}} 
\right)^2.
\end{align}
The function $\sigma^2$ is the usual variance estimator written in terms of $\bs n$.
\item For inference on the risk ratio
\be\label{thetarr}
\theta_{\mathrm{RR}} (\bs v ) = \frac{v_{11} + v_{10}}{v_{11} + v_{01}},
\ee
we may use the usual point estimator 
\be\label{trr}
T_{\mathrm{RR}}(\bs n) = \frac{n_{11} ( n_{01} + n_{00} )}{n_{01} (n_{11} + n_{10})}
\ee 
to construct the test statistic $R_{\mathrm{RR}}(\bs n,\bs v) = T_{\mathrm{RR}}(\bs n)  - \theta_{\mathrm{RR}} (\bs v )$.\footnote{In each of these expressions (and the ones in the next example), we interpret division by zero as returning some arbitrarily selected real number, such as $10^6$. This case is rare: as long as one treated unit has outcome 0 and one control unit has outcome 1, \eqref{thetarr} and \eqref{trr} will have nonzero denominators.}
\item For inference on the odds ratio 
\be
\theta_{\mathrm{OR}}(\bs v) = \frac{v_{11} + v_{10}} {v_{01} + v_{00}} \cdot \frac{v_{10} + v_{00}}{ v_{11} + v_{01}},
\ee
we may use the point estimate
\be
T_{\mathrm{OR}}(\bs n) = \frac{n_{11} n_{00}}{n_{10} n_{01}}
\ee 
to construct the test statistic $R_{\mathrm{OR}}(\bs n,\bs v) = T_{\mathrm{OR}}(\bs n)  - \theta_{\mathrm{OR}} (\bs v )$.
\end{itemize}

We now present our construction of confidence intervals for general estimands in the unbalanced case, which returns Monte Carlo approximations to $\mathcal I_\alpha$. The gain in efficiency relative to a direct search comes from the reuse of Monte Carlo samples implemented in Step 2(d), instead of generating new Monte Carlo draws to test each vector $\bs v$.  

\begin{alg}\label{a:unbalancedefficient}
This algorithm takes as input $\alpha \in (0,1)$, $K \in \mathbb{Z}_{>0}$,  a vector $\bs n \in \mathbb{Z}_{\ge 0}^{\{0,1\}^2}$, an estimand $\theta$, and a test statistic $R$. It returns a confidence set $\mathcal I_{\alpha,  K}$.\footnote{We suppress the dependence on the confidence set on $\theta$ and $R$ for brevity.}

We initialize $\mathcal I_{\alpha,  K} \leftarrow \varnothing$. For every $\tau_0 \in \mathcal C (\bs n)$, in descending order from $\max( \mathcal C (\bs n))$, we perform the following steps.

\begin{enumerate}
\item Initialize $j\leftarrow 0$.
\item\label{step22} If $j \le n$, perform the following steps.
\begin{enumerate}

\item\label{stepa2} 
Construct a count vector $\bs v$ as follows (recall \eqref{vdef}). 
Let $v_{10}$ be a free parameter, and set $v_{11} = j - v_{10}$. Then (solving \eqref{a1}) we set
\be\label{constructv2}
v_{01} = v_{10} - n \tau_0,\qquad 
v_{00} = n - j - v_{10} + n \tau_0.
\ee

\item \label{stepb2} 
Find the smallest $v_{10}$ that leads to a possible table $\bs v$ given the observed data $\bs n$. This can be done in constant time using  \Cref{l:possibleinequalities}.

If no value of $v_{10}$ leads to a possible table, set $j \leftarrow j +1$ and return to \eqref{step22}. 

\item\label{step52} Monte Carlo permutation test the count vector $\bs v$ coming from the choice of $v_{10}$ in the previous step using \Cref{a:approximatetest2} with the given values of $\alpha$, $K$, $\theta$, and $R$. If $\bs v$ is accepted, set $\mathcal I_{\alpha, K} \leftarrow \mathcal I_{\alpha, K} \cup \{ \theta(\bs v) \} $ using the returned value of $\theta(\bs v)$. 

\item\label{step6} 
Using the returned decision and $K$ samples $\bs{ Z}^{(i)}$ from the previous step, use the procedure described below in \Cref{s:step2d} to return a set $\mathcal J(\bs v) \subset \mathbb{R}$.
\item Set $\mathcal I_{\alpha, K} \leftarrow \mathcal I_{\alpha, K} \cup \mathcal J(\bs v) $. 

\end{enumerate} 
\end{enumerate}
\end{alg}
\begin{remark}
A small improvement in efficiency is possible for inference on the sample average treatment effect with $R = T - \tau$, since the second step may be terminated as soon as some value of $\tau_0$ is deemed compatible. For general statistics, all $\bs v$ with $\tau (\bs v) = \tau_0$ must be considered.
\end{remark}

\subsubsection{Step 2(d)}\label{s:step2d}
We now describe how to perform Step~2(d) of \Cref{a:unbalancedefficient}. Let $\bs v$ be the vector defined in \eqref{constructv2} using the choice of $v_{10}$ from Step 2(b).
For all $k \in \mathbb{Z}_{> 0}$, denote 
\be
\bs v_k = ( v_{11} - k, v_{10} + k , v_{01} +k, v_{00} -k)
\ee
and let
\be\label{theline}
\mathcal L = \{ \bs v_k : k \in \unn{1}{n}, \text{$\bs v_k$ is possible given $\bs n$} \}.
\ee
We will reuse the result of the approximate permutation test on $\bs v$ from Step 2(c) to permutation test each of the values in $\mathcal L$. The estimand values corresponding to the accepted values will form $\mathcal J$. 

The result of one of the $K$ random assignments $\{ \bs{ Z}^{(i)}\}_{i=1}^K$ returned by the previous step is summarized by a \emph{permutation vector} 
\be
\bs q = 
\big( 
q_{11}(0), q_{11}(1), q_{10}(0), q_{10}(1),
q_{01}(0), q_{01}(1), q_{00}(1), q_{00}(0)
\big),
\ee
where $q_{11}(0)$ denotes the number of $(1,1)$ vectors in the table $\tilde {\bs w}$ from \Cref{a:approximatetest2} that are assigned to control, $q_{11}(1)$ denotes the number of such columns assigned to treatment, and the rest are of the entries are described similarly.  Note that $q_{11}(0) + q_{11}(1) = v_{11}$, and analogous identities hold for $v_{10}$, $v_{01}$, and $v_{00}$. Also,
\be
q_{11}(1) +  q_{10}(1) + 
q_{01}(1) +  q_{00}(1)=  m
\ee
and similarly for the control group. Altogether, Step~2(c) yields a collection of $K$ permutation vectors $\{ \bs q^{(i)} \}_{i=1}^K$.


Given this set, we now  compute a collection of permutation vectors $\{ \bs {q'}^{(i)} \}_{i=1}^K$ for the count vector $\bs v' = \bs v  + (-1 ,1 ,1 -1)$. Given a permutation vector $\bs q$ for $\bs v$, we consider a Bernoulli random variable $B_1$ with 
\be
\P (B_1 = 1) = \frac{q_{00}(0)}{q_{00}(1) + q_{00}(0)}, \qquad \P (B_1 = 0 ) = \frac{q_{00}(1)}{q_{00}(1) + q_{00}(0)}.
\ee
If $B_1=1$, we set 
\be
q'_{00}(0) = q_{00}(0) -1, \qquad q'_{01}(0) = q_{01}(0) + 1.
\ee
If $B_1=0$, we set 
\be
q'_{00}(1) = q_{00}(1) -1, \qquad q'_{01}(1) = q_{01}(1) + 1
\ee
This has the effect of removing a potential outcome vector of the form $(0,0)$, chosen uniformly at random from all such vectors in $\bs w$, and replacing it with one of the form $(0,1)$.

In a similar way, we can exchange a potential outcome vector of the form $(1,1)$ for one equal to $(1,0)$.  Consider a random variable $B_2$ such that
\be
\P (B_2 = 1) = \frac{q_{11}(0)}{q_{11}(1) + q_{11}(0)}, \quad \P (B_2 = 0 ) = \frac{q_{11}(1)}{q_{11}(1) + q_{11}(0)}.
\ee
If $B_2=1$, we set 
\be
q'_{11}(0) = q_{11}(0) -1, \qquad q'_{10}(0) = q_{10}(0) + 1.
\ee
If $B_2=0$, we set 
\be
q'_{11}(1) = q_{11}(1) -1, \qquad q'_{10}(1) = q_{10}(1) + 1
\ee
The probabilities in each case are chosen in such a way that if $\bs q$ is induced by a uniform random assignment of subjects whose potential outcome table $\bs w$ corresponds to the count vector $\bs v$, then $\bs q'$ is distributed as if it were induced by a uniform random assignment of subjects whose potential outcome table $\bs w'$ corresponds to the count vector $\bs v'$. 

Given the collection of permutations vectors $\{ \bs q^{(i)} \}_{i=1}^K$ for $\bs v$, we  use the above recipe to obtain permutation vectors $\{ \bs q'^{(i)} \}_{i=1}^K$ for $\bs v'$ (distributed as if they were induced by a uniformly random assignment of a corresponding table $\bs w'$). Hence, we can compute an approximate $p$-value for $\bs v'$ as in \Cref{a:approximatetest2} using the $\{ \bs q'^{(i)} \}_{i=1}^K$. Iterating this procedure allows us to search the entire line segment $\mathcal L$ to find the set $\mathcal V$ consisting of $\bs v_k\in \mathcal L$ that are both possible and would be accepted by an application of \Cref{a:approximatetest2} (where possibility is checked using \Cref{l:possibleinequalities}). Finally, we let
\be
\mathcal J = \{ \theta(\bs v) : \bs v \in \mathcal V \}.
\ee

\subsection{Analysis}\label{s:IaKanalysis}

We now state on our main result on \Cref{a:unbalancedefficient}. The first part states that $\mathcal I_{\alpha, K}$ is a valid confidence set regardless of the value of $K$. The second bounds the running time. The proof is deferred to \Cref{a:unbalanced}. 
As in \eqref{e:shorthand}, we use $\mathcal I_{\alpha, K}( \bs Y , \bs Z)$ as shorthand for $\mathcal I_{\alpha , K}( \bs n)$, where $\bs n$ is the count vector for $(\bs Y, \bs Z)$. 

\begin{thm}\label{t:unbalanced}
Fix a potential outcome table $\bs y$, $\alpha \in (0,1)$, $K \in \mathbb{Z}_{ > 0}$, an estimand $\theta$, and a test statistic $R$. 
\begin{enumerate}
\item Let $\hat {\bs v}$ be the count vector for $\bs y$. We have 
\be\label{unbalancedclaim1}
\P\big ( \theta (\hat {\bs v}) \in \mathcal I_{\alpha,  K}(\bs Y , \bs Z) \big ) \ge 1 - \alpha.
\ee
\item The time complexity of \Cref{a:unbalancedefficient} with these parameters is $O( K n^3)$. 
\end{enumerate}
The probability statements are with respect to joint distribution of  $\bs Z$ and the random Monte Carlo draws $\bs Z^{(i)}$ used in each application of \Cref{a:approximatetest2}.
\end{thm}

Since a Monte Carlo permutation test with $K$ samples can be done in $O(Kn)$ time (by using Fisher--Yates algorithm to generate the $K$ permutations), one may think of the second part of \Cref{t:unbalanced} as saying that \Cref{a:unbalancedefficient} uses $O(n^2)$ ``effective'' approximate permutation tests, compared to the $O(n \log n)$ permutation tests of \Cref{a:approximate}.

\begin{remark}\label{remark:K}
We note that \Cref{t:unbalanced} holds for all values of $K$, so that the validity of the confidence interval procedure does not depend on $K$.\footnote{We are grateful to an anonymous referee for pointing out that this can be accomplished by adopting the definition in \eqref{approxpval2}.} However, larger values of $K$ generally lead to more powerful tests. Based on the simulations in \Cref{s:simulations}, we suggest choosing $K$ to be at least $10^4$. 

One naturally wonders whether a similar approach could be used in the balanced case, so that \Cref{a:approximate} gives valid confidence intervals regardless of the value of $K$. Such a modification does not seem straightforward due to the binary search used in that algorithm. For example, when searching for the upper bound, correctly returning $U_{\alpha, \eps, K}(\bs n)$ (or some larger value) requires that no value smaller  $U_{\alpha, \eps, K}(\bs n)$ is rejected during the search process. An incorrect rejection early in the search may lead to  $U_{\alpha, \eps, K}(\bs n)$ (or larger values) never being considered. For this reason, it seems that it is not enough  to ensure merely that each individual  permutation test gives a valid $p$-value, as in \Cref{r:pexact}
\end{remark}

\begin{remark}\label{r:both}
Both the balanced and unbalanced algorithms can be applied with the difference-in-means statistics. The time-complexity of the balanced algorithm is $O\left(n\times \textrm{polylog}(n)\right)$, by choosing $K=\log(4n\log^2n/\epsilon)/(2\epsilon^2)$  (following \eqref{K}), and for the unbalanced algorithm, the time complexity is $O(Kn^3)$. Hence, for large $n$, the balanced algorithm will be more efficient than the unbalanced algorithm.  However, the constant in the time complexity of the balanced algorithm is nontrivial, so in problems of small sizes, the unbalanced algorithm may be practically faster. This raises the question of what sample sizes each algorithm is best suited for. We discuss this point in \Cref{subsection:time} using simulations.
\end{remark}


\section{Missing Data}\label{s:missing}

We now consider a more general situation in which the realized outcomes $Y_i$ may not be observed for some subjects, with no restriction on the distribution of the unobserved $Y_i$. More precisely, we suppose that instead of observing $(\bs Y, \bs Z)$, the experimenter observes $(\bs M, \bs Z)$, where 
\be
\bs M \in \{-1,0,1\}^n,\qquad M_i = -J_i + (1 - J_i) Y_i,
\ee
and $\bs J \in \{1, 0 \}^{n}$. Here $\bs J$ is a random variable which may have an arbitrary distribution, and in particular may depend on $\bs Y$ and $\bs Z$. The indices $i$ such that $J_i = 0$ denote realized outcomes that are observed, while those such that $J_i = 1$ indicate missing data. Hence, we have $M_i = -1$ if the realized outcome for the $i$-th subject is not observed.

We can construct exact confidence intervals for $\tau(\bs y)$ given $(\bs M, \bs Z)$ by considering, roughly speaking, the two extremal imputations of the missing data (leading to the smallest and largest estimates of $\tau(\bs y)$).

\begin{definition}
Given a observed data $(\bs M, \bs Z)$, we define the vector $\bs Y^{(+)} \in \{0,1\}^n$ by
\begin{align*}
Y^{(+)}_i  &= 1 \text{ if } J_i =1 \text{ and } Z_i = 1,\\
Y^{(+)}_i  &= 0 \text{ if } J_i =1 \text{ and } Z_i = 0,\\
Y^{(+)}_i  &= M_i \text{ if } J_i =0.
\end{align*}
Similarly, we define $ \bs Y^{(-)}\in \{0,1\}^n$ by 
\begin{align*}
Y^{(-)}_i  &= 0 \text{ if } J_i =1 \text{ and } Z_i = 1,\\
Y^{(-)}_i  &= 1\text{ if } J_i =1 \text{ and } Z_i = 0,\\
Y^{(-)}_i  &= M_i \text{ if } J_i =0.
\end{align*}
Finally, for $\alpha \in (0,1)$, we set
\be
\mathcal I^\circ_\alpha(\bs M, \bs Z) = \left [
\min \big( L_\alpha(\bs Y^{(-)}, \bs Z),  T(\bs Y^{(-)}, \bs Z) \big), 
\max\big (U_\alpha(\bs Y^{(+)}, \bs Z), T(\bs Y^{(+)}, \bs Z) \big)
\right].
\ee
\end{definition}
The following proposition is proved in \Cref{s:missingproofs}. 
\begin{prop}\label{p:missing}
Suppose $m < n \in \mathbb{Z}_{>0}$, and fix potential outcomes $\bs y$ and $\alpha \in (0,1)$. Then 
\be
\P\big ( \tau(\bs y) \in \mathcal I^\circ_\alpha(\bs M, \bs Z) \big ) \ge 1 - \alpha,
\ee
where the probability is with respect to the variable $\bs Z$.
\end{prop}
\begin{remark}
In the balanced case $n = 2m$, \Cref{l:containsobserved} implies that the definition of
$\mathcal I^\circ_\alpha(\bs M, \bs Z)$ simplifies to 
\[
\mathcal I^\circ_\alpha(\bs M, \bs Z) = \big [ L_\alpha(\bs Y^{(-)}, \bs Z), U_\alpha(\bs Y^{(+)}, \bs Z)  \big].
\]
\end{remark}

\begin{remark}
When $n$ is odd, the intervals $\mathcal I_\alpha^\circ$ can be used to adapt \Cref{a:efficient}, or its Monte Carlo counterpart, in the following way. Given a group of $2m-1$ subjects, insert a fictitious subject to create a group of $2m$ subjects, randomize to equal groups, and construct $\mathcal I_\alpha^\circ$ by treating the data from the fictitious subject as missing. An argument similar to the one that proves \Cref{p:missing} shows that this interval covers the sample average treatment effect for the original group of $2m-1$ subjects with probability at least $1-\alpha$. While the method using $\mathcal I_\alpha^\circ$ will lose an $O(n^{-1})$ amount of precision relative to $\mathcal I_\alpha$, it permits the application of the faster binary search algorithm.
This trade-off between precision and computational practicality may favor using the $\mathcal I_\alpha^\circ$ construction when $n$ is large.
\end{remark}

\section{Simulations}\label{s:simulations}

\subsection{Comparison of Methods}
We begin by conducting three sets of simulations to evaluate our proposed methods. The first is in the balanced setting and compares the number of permutation tests required by \Cref{a:efficient} to the number required by previous proposals. The results, shown in \Cref{thetable}, are  improvements over the previous algorithms of Li--Ding and Rigdon--Hudgens \cite{rigdon2015randomization,li2016exact}.

\begin{table}[htb]
\centering
\begin{tabular}{|c|c|c|c|c|}
\hline
 $\bs n$ &  Confidence Interval & \Cref{a:efficient} & \cite{li2016exact}  & \cite{rigdon2015randomization} \\ \hline
 (2,6,8,0)& $[-14,-5]$ & 24 & 113 & 189 \\ 
 (6,4,4,6) & $[-4,10]$ & 16 & 308 &  1225 \\ 
 (8,4,5,7) & $[-3,13]$ & 26 & 421 & 2160 \\ \hline
\end{tabular}
\caption{95\% confidence intervals for $n\tau(\bs n)$. The second column gives the confidence interval $\mathcal I(\bs n)$, after scaling by $n$. The remaining columns indicate the number of permutation tests required to return $\mathcal I_\alpha$ for \Cref{a:efficient} and the algorithms from \cite{rigdon2015randomization,li2016exact}. Data for the last two columns is from \cite[Table I]{li2016exact}. 
}
\label{thetable}
\end{table}

The second set of simulations considers balanced trials with larger values of $n$. The results are reported in \Cref{table2}. We compare the confidence intervals (CIs) returned by \Cref{a:efficient}, based on permutation tests with the difference-in-means statistic, with asymptotic intervals based on the studentized difference-in-means statistic and a normal critical value. The third set considers unbalanced trials with treatment probability $3/5$ and is shown in \Cref{table3}. Here, we compare the CIs from \Cref{a:unbalancedefficient} with the difference-in-means statistic, CIs from \Cref{a:unbalancedefficient} with the studentized difference-in-means statistic, and asymptotic intervals based on the studentized difference-in-means statistic and a normal critical value.  

With $n=1000$ and a single thread on a contemporary processor, the balanced algorithm takes around $45$ minutes, and the unbalanced algorithm takes around $14.5$ hours. Given the ease of parallelizing our proposed algorithms, it is not hard to significantly improve these times by invoking multithreading.

\begin{table}[htb]
\centering
\begin{tabular}{|c|c|c|c|c|}
  \hline
   &\multicolumn{4}{|c|}{\textbf{First Scenario}} \\
  \hline
  &\multicolumn{2}{|c|}{Permutation} & 
  \multicolumn{2}{|c|}{Wald}\\
\hline

Sample Size & Cov.  & Med.\ Length  & Cov. & Med.\ Length  \\
\hline
50 & 0.97 & 0.37 & 0.93 & 0.39 \\ 
\hline
  100 & 0.97 & 0.37 & 0.93 & 0.39 \\ 
\hline
  200& 0.96 & 0.26 & 0.93 & 0.28 \\ 
  \hline
1000 &  0.96 & 0.12 & 0.95 & 0.12\\
\hline
  & \multicolumn{4}{|c|}{\textbf{Second  Scenario}} \\
  \hline
    &\multicolumn{2}{|c|}{Permutation} & 
  \multicolumn{2}{|c|}{Wald}\\
\hline
Sample Size & Cov.  & Med.\ Length  & Cov.  & Med.\ Length  \\
\hline
50 & 1.00 & 0.34 & 0.89 & 0.30 \\ 
\hline
  100 & 0.99 & 0.26 & 0.94 & 0.21 \\ 
\hline
  200 & 0.98 & 0.18 & 0.93 & 0.15 \\
   \hline
1000 &    0.99 & 0.08 & 0.96 & 0.07\\
\hline
\end{tabular}
\caption{Simulation results for the balanced case with sample sizes 50, 100, and 200. The numbers of treated units are 25, 50, and 100, respectively. In the first simulation scenario, the shares of units with outcomes $(1,1)$ and $(0,0)$ are 0.5. In the second simulation scenario, the share of units with outcome $(1,1)$ is 0.08 and the share of units with outcome $(0,0)$ is 0.92. The first column (Sample Size) reports sample sizes. The second column (Permutation Cov.) reports the coverage probability of the 95\% confidence intervals (CIs) based on the permutation test with the difference-in-means statistic. The third column (Permutation Med.\ Length) reports the median length of these CIs. 
The fourth column (Wald Cov.) reports the coverage probability of the 95\% CIs based on the studentized difference-in-means statistic and a normal critical value. The final column (Wald Med.\ Length) reports the corresponding median length of the CIs. The number of simulations is $10^4$, and the number of randomizations is chosen according to the formula \eqref{K} with $\epsilon=0.005$. All numbers are rounded two digits after the decimal point. 
}
\label{table2}
\end{table}


Tables 2 and 3 consider two simulation scenarios.
The first scenario is a case where the shares of units with outcomes $\left(y(1),y(0)\right)=\left(1,1\right)$ and $\left(y(1),y(0)\right)=\left(0,0\right)$ are each $0.5$. The second scenario is a sparse-outcome case where the share of units with outcome $(1,1)$ is $0.08$ and the share of units with outcome $(0,0)$ is $0.92$. Our parameter of interest is the sample average treatment effect, which is 0 by construction in all cases.
For both scenarios, we report coverage probabilities and median CI length for the sample sizes 50, 100, and 200. We also report results in the balanced case with a sample size of 1000. 
We observe that the asymptotic CIs may exhibit significant undercoverage. For example, for $n=50$ in the balanced case with sparse data, the coverage probability is $89\%$ for a nominally $95\%$ confidence interval. In contrast, the CIs based on our proposed methods provide the nominal coverage, and the length of the CIs is comparable to those based on the Wald test (and sometimes shorter).

In binary outcome settings, confidence intervals based on Wald statistics are known to undercover when the data are sparse (i.e., when there are few 1's or few 0's) \cite{brown2001interval}. Although the Wald-based interval is asymptotically valid for a fixed 
 proportion of $0$ and $1$ outcomes, it is unclear how large the sample size must be for the asymptotic approximation to hold, especially since the true proportion is unknown to the analyst.
When the asymptotics are not reliable, Wald intervals risk undercoverage, and our methods become preferable. 
Examples of potentially sparse datasets include those arising in online experiments (when few users respond or baseline activity is low) and in health studies (when adverse outcomes are rare). Whenever computationally feasible, our procedure provides an attractive alternative in such settings.

\begin{table}[htb]
\centering
\begin{tabular}{|c|c|c|c|c|c|c|}
  \hline
   &\multicolumn{6}{|c|}{\textbf{First  Scenario}} \\
  \hline
& \multicolumn{2}{|c|}{Permutation (DiM)} & \multicolumn{2}{|c|}{Permutation (Std)} & \multicolumn{2}{|c|}{Wald}\\
\hline
Sample Size & Cov.  & Med.\ Length  & Cov. & Med.\ Length  & Cov.  & Med.\ Length  \\
\hline  
50 & 0.96 & 0.48 & 0.96 & 0.52 & 0.96 & 0.57 \\ 
\hline
 100 & 0.95 & 0.36 & 0.97 & 0.37 & 0.93 & 0.40 \\
\hline
 200 & 0.96 & 0.27 & 0.96 & 0.27 & 0.94 & 0.28 \\ 
\hline
&\multicolumn{6}{|c|}{\textbf{Second  Scenario}} \\
  \hline
& \multicolumn{2}{|c|}{Permutation (DiM)} & \multicolumn{2}{|c|}{Permutation (Std)} & \multicolumn{2}{|c|}{Wald}\\
\hline
Sample Size & Cov.  & Med.\ Length  & Cov. & Med.\ Length  & Cov.  & Med.\ Length  \\
\hline
50 & 0.98 & 0.34 & 0.98 & 0.28 & 0.86 & 0.32 \\ 
\hline
  100 & 0.98 & 0.25 & 0.98 & 0.19 & 0.98 & 0.22 \\ 
\hline
  200 & 0.98 & 0.18 & 0.97 & 0.14 & 0.93 & 0.15 \\ 
\hline
\end{tabular}
\caption{Simulation results for the unbalanced case with sample sizes 50, 100, and 200. The numbers of treated units are 30, 60, and 120, respectively. In the first simulation scenario, the shares of units with outcomes $(1,1)$ and $(0,0)$ are 0.5. In the second simulation scenario, the share of units with outcome $(1,1)$ is 0.08 and the share of units with outcome $(0,0)$ is 0.92. The first column (Sample Size) reports sample sizes. The second column (Permutation (DiM) Cov.) reports the coverage probability of the 95\% CIs based on the permutation test with the difference-in-means statistic. The third column (Permutation (DiM) Med.\ Length) reports the median length of the 95\% CIs based on the permutation test with the difference-in-means statistic. The fourth column (Permutation (Std) Cov.) reports the coverage probability of the 95\% CIs based on the studentized difference-in-means statistic. The fifth column (Permutation (Std) Med.\ Length) reports the corresponding median length of the CIs. The sixth column (Wald Cov.) reports the coverage probability of the 95\% CIs based on the studentized difference-in-means statistic and a normal critical value. The seventh column (Wald Med.\ Length) reports the corresponding median length of the CIs. The number of simulation is $10^4$, and the number of randomizations is $2\times 10^4$.  All numbers are rounded two digits after the decimal point.}
\label{table3}
\end{table}

\subsection{Computation Time}\label{subsection:time}
As noted in Remark \ref{r:both}, both the balanced and unbalanced algorithms can be applied to the difference-in-means statistic. We evaluate their run times using the same simulation setups as above. Tables \ref{table4} and \ref{table5} report the average runtimes for simulation scenarios 1 and 2, respectively. For the balanced algorithm, the number of randomizations is determined by the formula \eqref{K} with $\epsilon=0.005$. For the unbalanced algorithm, we consider $K \in \{1000, 2000, 5000, 10000, 20000\}$. Tables \ref{table6} and \ref{table7} present the widths of the confidence intervals produced by each algorithm. 

We highlight two key observations from the results.  
First, as shown in Tables \ref{table6} and \ref{table7}, for the unbalanced algorithms, the lengths of the confidence intervals decrease as $K$ increases, providing supporting evidence for Remark \ref{remark:K}. Second, in Tables \ref{table4} and \ref{table5}, for $n = 50$, $100$, and $200$, the unbalanced algorithm is practically faster than the balanced algorithm. However, when $n = 500$ or $1000$ and $K \geq 5000$, the balanced algorithm typically outperforms the unbalanced one. 
We recommend that readers use the unbalanced algorithms when the sample size is small and the balanced algorithms when the sample size is larger, with a threshold informed by our simulation results (in the low hundreds).
\begin{table}[ht]
\centering
\begin{tabular}{|c|c|ccccc|}
  \hline
     &\multicolumn{6}{|c|}{\textbf{First  Scenario}} \\
     \hline
 Sample Size & Balanced & \multicolumn{5}{c|}{Unbalanced} \\
  \hline
 &  & $K=1$,000 & $K=2$,000 & $K=5$,000 & $K=1$0,000 & $K=2$0,000 \\ 
  \hline
50 & 13.23 & 0.23 & 0.46 & 1.14 & 2.37 & 4.85 \\ 
100& 63.59 & 2.33 & 4.75 & 12.09 & 25.17 & 52.20 \\ 
200 & 249.15 & 20.41 & 41.97 & 107.71 & 226.55 & 497.06 \\ 
500 & 809.50 & 340.62 & 642.32 & 1720.28 & 4024.48 & 8493.88 \\ 
1000 & 2270.71 & 2354.81 & 4713.48 & 13461.07 & 29531.78 & 62791.92 \\ 
   \hline
\end{tabular}
\caption{Average computation times (in seconds) for the balanced algorithm and unbalanced algorithm with the difference-in-means statistic and sample sizes 50, 100, 200, 500, and 1000 in Simulation Scenario 1. The numbers of treated units are 25, 50, 100, 250, and 500, respectively. The number of simulations is $10^4$ for sample sizes 50, 100, and 200, and $10^3$ for sample sizes 500 and 1000. The number of randomizations for the balanced algorithm is chosen according to the formula \eqref{K} with $\epsilon=0.005$. All numbers are rounded to two digits after the decimal point.}
\label{table4}
\end{table}

\begin{table}[ht]
\centering
\begin{tabular}{|c|c|ccccc|}
  \hline
       &\multicolumn{6}{|c|}{\textbf{Second  Scenario}} \\
     \hline
  Sample Size & Balanced & \multicolumn{5}{c|}{Unbalanced} \\
  \hline
 &  & $K=1$,000 & $K=2$,000 & $K=5$,000 & $K=1$0,000 & $K=2$0,000 \\ 
  \hline
50 & 14.08 & 0.24 & 0.47 & 1.14 & 2.31 & 4.71 \\ 
 100 & 78.20 & 1.39 & 2.81 & 7.05 & 14.41 & 29.40 \\ 
200 & 215.11 & 11.98 & 24.29 & 61.24 & 125.94 & 258.15 \\ 
  500 & 1182.66 & 439.95 & 282.34 & 439.95 & 1590.83 & 3247.27 \\ 
  1000 & 2850.90 & 833.64 & 1657.10 & 4313.17 & 9819.47 & 21568.74 \\ 
   \hline
\end{tabular}
\caption{Average computation times (in seconds) for the balanced algorithm and unbalanced algorithm with the difference-in-means statistic and sample sizes 50, 100, 200, 500, and 1000 in Simulation Scenario 2. The numbers of treated units are 25, 50, 100, 250, and 500, respectively. The number of simulations is $10^4$ for sample sizes 50, 100, and 200, and $10^3$ for sample sizes 500 and 1000. The number of randomizations for the balanced algorithm is chosen according to the formula \eqref{K} with $\epsilon=0.005$. All numbers are rounded to two digits after the decimal point.}
\label{table5}
\end{table}
\begin{table}[ht]
\centering
\begin{tabular}{|c|c|ccccc|}
  \hline
       &\multicolumn{6}{|c|}{\textbf{First  Scenario}} \\
     \hline Sample Size
  & Balanced & \multicolumn{5}{c|}{Unbalanced} \\
  \hline
 &  & $K=1$,000 & $K=2$,000 & $K=5$,000 & $K=10$,000 & $K=20$,000 \\ 
 \hline
50 & 0.50 & 0.49 & 0.48 & 0.48 & 0.48 & 0.48 \\ 
100 & 0.37 & 0.37 & 0.36 & 0.36 & 0.36 & 0.36 \\ 
200 & 0.27 & 0.27 & 0.27 & 0.26 & 0.26 & 0.26 \\ 
500 & 0.17 & 0.18 & 0.18 & 0.17 & 0.17 & 0.17 \\ 
1000 & 0.12 & 0.13 & 0.13 & 0.12 & 0.12 & 0.12 \\ 
   \hline
\end{tabular}
\caption{Average widths for the balanced algorithm and unbalanced algorithm with the difference-in-means statistic and sample sizes 50, 100, 200, 500, and 1000 in Simulation Scenario 1. The numbers of treated units are 25, 50, 100, 250, and 500, respectively. The number of simulations is $10^4$ for sample sizes 50, 100, and 200, and $10^3$ for sample sizes 500 and 1000. The number of randomizations for the balanced algorithm is chosen according to the formula \eqref{K} with $\epsilon=0.005$. All numbers are rounded to two digits after the decimal point.}
\label{table6}
\end{table}

\begin{table}[ht]
\centering
\begin{tabular}{|c|c|ccccc|}
  \hline
       &\multicolumn{6}{|c|}{\textbf{Second  Scenario}} \\
     \hline Sample Size
  & Balanced & \multicolumn{5}{c|}{Unbalanced} \\
  \hline
 &  & $K=1$,000 & $K=2$,000 & $K=5$,000 & $K=10$,000 & $K=20$,000 \\ 
  \hline
50 & 0.35 & 0.35 & 0.35 & 0.35 & 0.35 & 0.35 \\ 
100 & 0.26 & 0.29 & 0.29 & 0.28 & 0.28 & 0.28 \\ 
200& 0.18 & 0.19 & 0.18 & 0.18 & 0.18 & 0.18 \\ 
500& 0.12 & 0.12 & 0.12 & 0.12 & 0.11 & 0.11 \\ 
1000 & 0.08 & 0.09 & 0.08 & 0.08 & 0.08 & 0.08 \\ 
   \hline
\end{tabular}
\caption{Average widths for the balanced algorithm and unbalanced algorithm with with the difference-in-means statistic and sample sizes 50, 100, 200, 500, and 1000 in Simulation Scenario 1. The numbers of treated units are 25, 50, 100, 250, and 500, respectively. The number of simulations is $10^4$ for sample sizes 50, 100, and 200, and $10^3$ for sample sizes 500 and 1000. The number of randomizations for the balanced algorithm is chosen according to the formula \eqref{K} with $\epsilon=0.005$. All numbers are rounded to two digits after the decimal point.}
\label{table7}
\end{table}

\section{Open Problems}\label{s:open}

Many  questions remain. First, it is natural to ask whether an even more efficient construction of $\mathcal I_\alpha$ is possible, either by decreasing the number of permutation tests necessary or finding a more efficient way to compute (approximate) permutation $p$-values. We are unaware of any formal analysis of the computational hardness of such questions. Regarding the Monte Carlo permutation tests, we do not know how to implement any of the standard variance reduction techniques in this context, such as importance sampling, which would provide a practical (though non-asymptotic) improvement in run time. These methods have been successfully applied to permutation inference in other settings \cite{branson2019randomization,mehta1988importance}.

Second, a number of questions arise from relaxing the various assumptions made earlier. For example, is efficient inference possible in the case of a general bounded and discrete outcome, or a continuous outcome supported on a bounded interval? For a discussion of these questions in the context of estimating quantiles of individual treatment effects, see \cite{caughey2021randomization}.

Third, it is natural to ask about the true asymptotic behavior of the permutation intervals (as opposed to the crude bound in \Cref{rate}). The results in \cite{chung2013exact}, obtained in the superpopulation context, strongly suggest that in an appropriate $n\rightarrow \infty$ limit of experiments, the attained coverage will asymptotically approach $1-\alpha$ in both the balanced case (with the difference-in-means statistic) and the unbalanced case (with the studentized test statistic). However, this prediction remains to be rigorously verified.

Fourth  a more realistic setup is to suppose that each subject comes with additional data, such as demographic information and medical history, that is correlated with the potential outcomes. Then it desirable to find ways to leverage to this covariate information to increase precision. In this context, stratified designs, where subjects are independently randomized to treatment or control within subgroups formed using these additional covariates, are a natural choice.  

Rigdon and Hudgens show that their algorithm adapts to designs using stratified randomization,  \cite{rigdon2015randomization}. However, they observe that a straightforward adaptation of their approach quickly becomes computationally intractable, requiring $O( n^{4k})$ permutation tests, where $k$ is the number of strata. 

To understand the difficulty that strata create, suppose that each subject comes equipped with a binary covariate $X$, and we conduct complete randomization separately among the subjects with $X=0$ and the subjects with $X=1$. 
Then the possible potential outcome tables now take values in $\mathbb{Z}_{\ge 0}^{\{0,1\}^3}$ instead of $\mathbb{Z}_{\ge 0}^{\{0,1\}^2}$, with an additional degree of freedom coming from the new binary covariate, increasing the number of potential outcome count vectors $\bs v$ to examine from $O(n^3)$ to $O(n^7)$.\footnote{Without the covariate, there are $4$ coordinates to choose from, with the single linear restriction that they sum to $n$. With the binary covariate, there are $8$ coordinates and a single linear restriction.} 

In \Cref{s:general}, when we examined the case without covariates, we saw that a Monte Carlo algorithm that tests each of these $\bs v$ and re-used Monte Carlo draws could achieve a running time equivalent to $O(n^2)$ approximate permutation tests, a factor of $n$ gain over the naive approach. One can similarly gain a factor of $n$ when there is a binary covariate, but this only brings the time complexity to $O(n^6)$, which is far from practical.  It would be interesting to find more efficient methods to analyze stratified designs. Once found, these methods could likely also be adapted to construct covariate-adjusted confidence intervals using poststratification estimators \cite{miratrix2013adjusting,branson2019randomization2,pashley2021conditional}. 

Finally, it is of practical importance to find an efficient algorithm for prospective sample size calculations for the exact intervals $\mathcal I_\alpha$. Researchers often ask how many experimental subjects are needed to detect an effect with high probability (that is, have $0 \notin \mathcal I_\alpha$), assuming the effect is larger than some given threshold. If too few subjects are used, the experiment is uninformative, and if too many subjects are used, the experiment is potentially wasteful. For the Rigdon--Hudgens intervals, it is natural to estimate power through simulations using various plausible configurations of potential outcomes. However, it would be desirable to have a more efficient and rigorous technique.

\section{Acknowledgments}
Patrick Lopatto was supported by NSF postdoctoral
fellowship DMS-220289. The authors are grateful to the three anonymous referees for their detailed and thoughtful suggestions, which greatly improved the paper.
\bibliographystyle{alpha}
\bibliography{randomization}

\appendix

\section{Proofs for Section~\ref{s:basic}}\label{s:basicproofs}

The following definition will be used in the proofs of our results. For $k \in \mathbb Z_{> 0}$, we denote
\be
k^{-1} \mathbb Z  =  \left\{ \frac{j}{k} : j \in \mathbb{Z} \right\}, \quad k^{-1} (2\mathbb Z)  =  \left\{ \frac{2j}{k} : j \in \mathbb{Z} \right\},
\quad k^{-1} (2\mathbb Z + 1)  =  \left\{ \frac{2j +1}{k} : j \in \mathbb{Z} \right\}.
\ee

\begin{definition}
Let $f \colon \mathbb{Z} \rightarrow  \bbR_{\ge 0}$ be a probability mass function. We say that $f$ is \emph{symmetric about $k_0$} for some $k_0 \in 2^{-1} \mathbb Z$ if 
\be
f(k_0 - x) = f(k_0 + x) \text{ for all $x\in 2^{-1} \mathbb Z$ such that $k_0 + x \in \mathbb{Z}$.}
\ee
If $f$ is symmetric about $k_0$, we say that $f$ is \emph{decreasing away from $k_0$} if 
\be
f(k_0 + y) \le f(k_0+x)\text{ for all $x,y \in 2^{-1} \mathbb Z$ such that $y >x$ and $k_0 + x ,k_0 + y \in \mathbb{Z}$.}
\ee 

We now suppose that $\sum_{k=-\infty}^\infty k f(k)$ converges absolutely, so that the mean of the distribution corresponding to $f$ exists. In this case, we say that $f$ is \emph{symmetric--decreasing} (abbreviated SD) if $f$ is symmetric about $k_0 = \sum_{k=-\infty}^\infty k f(k)$, and decreasing away from $k_0$.

Given $k \in \mathbb{Z}_{>0}$ and $k_0 \in 2^{-1} \mathbb Z$, we say that a probability mass function $f \colon k^{-1} \mathbb{Z} \rightarrow  \bbR_{\ge 0}$ is symmetric about $k^{-1} k_0 \in (2k)^{-1}\mathbb{Z} $ if the function $g\colon \mathbb{Z} \rightarrow \bbR_{\ge 0}$ defined by $g(x) = f(k x)$ is symmetric about $k_0$. Similarly, we say that $f$ is SD if $g(x) = f(kx)$ is.
\end{definition}

Recall from \eqref{e:wdef} that $\bs w$ denotes a generic potential outcome table. We use the notation $T(\bs w, \bs Z)$ to stand for $T(\bs Y, \bs Z)$ with $Y$ generated using the potential outcomes $\bs w$ and the treatment assignments given by $\bs Z$. Since this quantity depends on $\bs w$ only through its associated count vector $\bs v$, we also write $T(\bs v, \bs Z)$ for $T(\bs n, \bs Z)$. 

\subsection{Proof of Lemma~\ref{l:containsobserved}}

We begin with two preliminary lemmas. 

\begin{lem}\label{l:symmetric}
Let $n = 2m$ for some $m \in \mathbb{Z}_{>0}$. 
For any potential outcome table $\bs w$, the pmf of $T(\bs w, \bs Z)$ is symmetric about $\E\big[ T(\bs w, \bs Z)\big]$.
\end{lem}
\begin{proof}
Let $\bs Z^\dagger$ be the random vector defined by $Z^\dagger_i = 1 - Z_i$.
Then for any randomization given by $\bs Z$, the vector $\bs Z^\dagger$ represents a randomization that exchanges the control and treatment groups. We have 
\be\label{apple}
\frac{1}{2} \left( T(\bs w,  \bs Z) + T(\bs w,  \bs Z^\dagger) \right)
= \frac{1}{2m} \sum_{i=1}^n \big( w_j(1) - w_j(0) \big) =
\E\big[ T(\bs w,  \bs Z) \big].
\ee
Let $f$ be the pmf of $T(\bs w, \bs Z)$, which is supported on $m^{-1} \mathbb{Z}$. Since $\bs Z$ and $\bs Z^\dagger$ have the same distribution, we deduce from \eqref{apple} that 
\be
f\Big(\E\big[ T(\bs w,  \bs Z) \big] - j\Big) = f\Big(\E\big[ T(\bs w,  \bs Z) \big] + j\Big)
\ee
for all $j \in n^{-1} \mathbb{Z}$ such that $ \E \big[T(\bs w,  \bs Z) \big] + j \in m^{-1} \mathbb{Z}$, which completes the proof.
\end{proof}

\begin{lem}\label{l:containsobs}
Let $n = 2m$ for some $m \in \mathbb{Z}_{> 0}$, and fix observed data $\bs Y, \bs Z$. Then then there exists a potential outcome table $\bs w$ such that $\tau(\bs w) = T(\bs Y, \bs Z)$ and $\bs w$ is possible given $\bs Y, \bs Z$.
\end{lem}
\begin{proof}
We have $T(\bs Y, \bs Z) = m^{-1} (n_{11} - n_{01})$. Let $\bs w$ be a potential outcome table with count vector $\bs v = (0, n_{11} + n_{00}, n_{10} + n_{01}, 0 )$, and such that $\bs w$ is possible given $\bs Y, \bs Z$. This can be arranged by imputing the unknown potential outcomes so that all subjects have potential outcomes $(0,1)$ or $(1,0)$. Then using
\be
n_{11} + n_{10} = n_{01} + n_{00} = m,
\ee
we compute
\begin{align}
\tau(\bs w) &= \frac{1}{n} ( n_{11} + n_{00} -n_{10} - n_{01})\\
 &= 
\frac{1}{n} \big( n_{11} + (m - n_{01})  -( m - n_{11}) - n_{01}\big)
= \frac{2 n_{11} - 2n_{01}}{n} = T(\bs Y, \bs Z).
\end{align}
This completes the proof.
\end{proof}

\begin{proof}[Proof of \Cref{l:containsobserved}]
Let $\bs w$ be the potential outcome table provided by \Cref{l:containsobs}, so that $\tau(\bs w) = T(\bs Y, \bs Z)$ and $\mathcal I_\alpha(\bs n)  = \mathcal I_\alpha(\bs Y, \bs Z)$. Then the distribution of $T(\bs w, \bs Z)$ is symmetric about $T(\bs Y, \bs Z)$, so $T(\bs Y, \bs Z)$ is the median of the distribution. Then by \eqref{pvalue}, we have $p(\bs w, \bs Y, \bs Z) = 1$, which shows that $\tau( \bs w)$ will always be accepted in  \Cref{a:rh}. This completes the proof.
\end{proof}

\subsection{Proof of Proposition~\ref{p:coverage}}

\begin{proof}[Proof of \Cref{p:coverage}]
The event $\{ \tau(\bs y) \notin \mathcal I_\alpha(\bs Y , \bs Z)  \}$ is contained in the event where \Cref{a:rh} rejects $\tau(\bs y)$, which is in turn contained in the event where
\be\label{evt}
 \P\left( \big| T(\bs{\tilde Y}, \bs{ \tilde Z} )  - \tau(\bs y) \big| \ge \big| T(\bs Y, \bs Z )  - \tau(\bs y) \big| \right) < \alpha.
\ee
Here probability is with respect to $\bs{ \tilde Z}$, and we consider the left side of \eqref{evt} as a function of $T(\bs Y, \bs Z)$. Let $x$ denote the smallest real number such that  
\be
 \P\left( \big| T(\bs{\tilde Y}, \bs{ \tilde Z} )  - \tau(\bs y) \big| \ge x \right) < \alpha,
\ee
which exists because $T(\bs{\tilde Y}, \bs{ \tilde Z} )$ is a discrete random variable supported on a bounded set. Then the event where \eqref{evt} occurs is the same as the event where $\big| T(\bs Y, \bs Z )  - \tau(\bs y) \big|\ge x$. By the  definition of $x$, and the fact that $(\bs Y, \bs Z)$ and $(\bs{\tilde Y}, \bs{ \tilde Z} )$ are identically distributed, this event has probability at most $\alpha$. This completes the proof.
\end{proof}

\subsection{Proof of Proposition~\ref{rate}}

\begin{proof}[Proof of \Cref{rate}]
Let $\bs w$ be an arbitrary potential outcome table. Recall from \Cref{d:permtest}
that $\tilde{ \bs Y}$ is defined as in \eqref{bigY} using the potential outcomes $\bs w$ in place of $\bs y$, and $\tilde{\bs Z}$ is a vector with the same distribution as $\bs Z$. 
By \cite[Equation (6.3)]{imbens2015causal}, 
\be
\var\big( T(\tilde{ \bs Y}, \bs {\tilde  Z}) \big)  \le 
\frac{S_1^2}{m} + 
\frac{S_0^2}{n-m},
\ee
where
\begin{align}
S_1^2 &= \frac{1}{n-1} \sum_{i=1}^n \big( 
w_i(1)  - \overline{w}(1)
\big)^2, \qquad   \overline{w}(1) = \frac{1}{n} \sum_{i=1}^n w_i(1) \\
S_0^2 &= \frac{1}{n-1} \sum_{i=1}^n \big( 
w_i(0)  - \overline{w}(0)
\big)^2, 
\qquad
 \overline{w}(0) = \frac{1}{n} \sum_{i=1}^n w_i(0) .
\end{align}
Since the $w_i(j)$ are each $0$ or $1$, we have 
\be
S_1^2  \le \frac{n}{n-1}, \qquad S_0^2  \le \frac{n}{n-1}
\ee 
and hence 
\be\label{a15}
\var\big( T(\tilde{ \bs Y}, \bs {\tilde Z}) \big)  \le  \frac{n^2 }{n-1} \cdot \frac{1}{ m (n-m)}.
\ee 
Let $V$ denote the quantity on the right-hand side of \eqref{a15}. Then Chebyshev's inequality implies that
\be\label{cheby}
\P\left( 
 \big| T(\tilde{ \bs Y}, \bs {\tilde Z}) - \tau( \bs w) \big| \ge t 
\right) 
\le \frac{V }{t^2} ,
\ee 
where we used that $T(\tilde{ \bs Y}, \bs {\tilde Z})$ has mean $\tau( \bs w)$.

We introduce the shorthand
\be
\delta = \big| T({ \bs Y}, \bs { Z})  -\tau(\bs w)  \big|.
\ee
To reject $\bs w$, we must have 
\be
\P\Big(\big| T(\tilde{ \bs Y}, \bs {\tilde Z})  -\tau(\bs w)  \big|  \ge 
\delta 
 \Big)  < \alpha.
\ee
Using \eqref{cheby}, we note that $V \delta^{-2} < \alpha$ is a sufficient condition for rejection. We solve $V \delta^{-2} < \alpha$ to find 
\be
\delta > \sqrt{ \frac{V}{\alpha}}.
\ee 
Therefore the permutation test will reject all values of $\tau(\bs w)$ with distance greater than $V^{1/2} \alpha^{-1/2}$ from $T({ \bs Y}, \bs Z)$. This implies the conclusion.
\end{proof}

\section{Proofs for Section~\ref{s:efficient}}\label{s:efficientproofs}

\subsection{Preliminary Results}\label{s:eprelim}

For completeness, and to facilitate the proofs of our finite-sample bounds, we begin by specifying the version of binary search that we use. The binary search algorithm below searches for the right boundary of the interval. The binary search algorithm for the left boundary is similar and we omit the details. 

\begin{alg}\label{a:binary}
This \emph{binary search algorithm} takes as input $k_1, k_2\in \mathbb{Z}_{>0}$  such that $k_2 > k_1$, and a function $f\colon \unn{k_1}{k_2} \rightarrow \{0,1\}$ such that $f(x) = 0$ if $x \le r$ and $f(x) = 1$ if $x >r$, where $r \in \unn{k_1-1}{k_2}$ is unknown. The algorithm returns $r$ through the following steps.
\begin{enumerate}
\item Initialize $a\leftarrow k_1$ and $b \leftarrow k_2$. 
\item Set $c \leftarrow \lfloor (a + b )/2 \rfloor$ and evaluate $f(c)$. If $f(c) =0$, set $a \leftarrow c$. If $f(c) = 1$, set $b\leftarrow c$.
\item Repeat the previous step until $b = a + 1$. 
In this case, return $a$ if $a>k_1$ and $b<k_2$. If $a=k_1$, evaluate $f(k_1)$ and return $k_1-1$ if $f(k_1) =1$, and $k_1$ if $f(k_1)=0$. If $b=k_2$, evaluate $f(k_2)$, and return $k_2$ if $f(k_2) = 0$ and $k_2 - 1$ if $f(k_2)= 1$. 
\end{enumerate} 
\end{alg}

The following fact is well known. 

\begin{lem}\label{l:binarylog}
\Cref{a:binary} terminates in at most $\lfloor \log_2( k_2 - k_1 + 1 ) + 2  \rfloor$ evaluations of $f$. 
\end{lem}
\begin{proof}
Without loss of generality, we may suppose that $k_1 = 1$. Further, by setting $f(x) = 1$ for $ x > k _2$, we may consider $f$ as a function on $\unn{1}{2^d+1}$, where $d = \lfloor \log_2(k_2) + 1 \rfloor $ satisfies $2^d \ge k_2$. Then the number of integers in the interval $[a,b]$ after each evaluation of $f$ follows the sequence $2^{d-1} + 1,\dots, 2^1 + 1 ,2 $ as the algorithm progresses, with a possible additional evaluation of $f$ if $a=1$ or $b=2^d + 1$ at the final step. Then at at worst $d+1$ evaluations are needed in total, which proves the theorem.
\end{proof}

The following lemma determines the possible potential outcome tables, given some observed data, in a certain one-parameter family of tables.  We remark that it shows Step \eqref{stepb} of \Cref{a:efficient} can be completed in constant time. We also remark that it does not assume that the experiment is balanced.
\begin{lem}\label{l:possibleinequalities}
Fix $j, v_{10} \in \mathbb{Z}_{\ge 0}$ and $\tau_0 \in \mathcal C$, and consider the potential outcome vector 
\be
\bs v = (j - v_{10},
v_{10}, 
v_{10} - n \tau_0,
n - j  -  v_{10} + n \tau_0).
\ee
Given observed data $\bs n$, a necessary condition for $\bs v$ to be possible is that all of the following inequalities hold:
\be
j \ge n \tau_0 + n_{01}, \quad j \ge n_{11},
\quad n \ge j + n_{10}, \quad n_{11} + n\tau_0 + n_{10} + n_{01} \ge j.
\ee
In this case, the possible values of $v_{10}$ are given by the interval
\be
 \max(0, n \tau_0, j - n_{11} - n_{01}, n_{11} + n_{01} + n\tau_0 - j )
  \le v_{10} \le \min(j, n_{11} + n_{00} ,    n_{10} + n_{01} + n \tau_0, n+ n\tau_0 -j),
\ee\label{interval}
which may be empty. 
\end{lem}
\begin{proof}[Proof of \Cref{l:possibleinequalities}]
We apply the criterion of \Cref{l:ispossible}. The maximum in that statement becomes
\be
\max(0, n_{11} - v_{10}, j - v_{10} - n_{01}, j - n \tau_0 - n_{10} - n_{01}).
\ee
Similarly, the minimum becomes
\be
\min(j - v_{10}, n_{11}, j - n \tau_0  - n_{01}, n - v_{10}  -n_{10} - n_{01} ).
\ee
Each argument of the maximum function must be less than each argument of the minimum function. We test the arguments of the maximum from left to right. Starting with $0$, we must have
\be
j \ge v_{10}, \quad n_{11} \ge 0,\quad
j \ge n \tau_0 + n_{01},
\quad 
n - n_{01} - n_{10} \ge v_{10},
\ee
where we note that the condition $n_{11} \ge 0$ is always true.
Considering $n_{11} - v_{10}$,  we get
\be
j \ge n_{11}, \quad v_{10}\ge 0, \quad  v_{10} \ge n_{11} + n_{01} + n \tau_0 - j, n \ge n_{11} + n_{10} + n_{01}.
\ee
Note that the last condition is always satisfied. 
Considering $j - v_{10} - n_{01}$, we get
\be
n_{10} \ge 0, \quad v_{10}   \ge  j - n_{11} - n_{01} , \quad v_{10} \ge n \tau_0, \quad n \ge j + n_{10},
\ee
and the first condition is always true. 
Finally, considering $j - n\tau_0 - n_{10} - n_{01}$, 
\be
n\tau_0 + n_{10} + n_{01} \ge v_{10},
\quad
n_{11} + n\tau_0 + n_{10} + n_{01} \ge j,
\quad
n_{10} \ge 0, \quad 
n + n \tau_0 -j \ge v_{10}.
\ee
Collecting the previous four displays completes the proof.
\end{proof}

\subsection{Proof of Theorem~\ref{t:main}}\label{s:mainproof}
We now prove \Cref{t:main} assuming \Cref{l:monotonic}, which is proved below.
\begin{proof}[Proof of \Cref{t:main}]
We will show that \Cref{a:efficient} returns $\mathcal I_\alpha$ in the required number of permutation tests. It suffices to show that \Cref{a:efficient} returns $U^{(1)}_\alpha(\bs n) = U_\alpha(\bs n)$ for the upper bound in $(n+1)\lfloor \log_2 (n +1) + 3 \rfloor $ permutation tests, since the same estimate holds for the lower bound $L_\alpha$ by analogous reasoning. Then the conclusion follows after  using 
\be 4 n \log_2(n) \ge 2(n+1)\lfloor \log_2 (n +1) + 3 \rfloor\ee
for $n \ge 15$.

 Let $C_+ = \max\big(\mathcal C(\bs n)\big)$, where we recall that $\mathcal C(\bs n)$ was defined in \eqref{mathcalC}. By \Cref{l:containsobserved}, we have $U_\alpha(\bs n) \in \big[ T(\bs n ), C_+ \big]$. Recall that by  \Cref{l:isinterval}, the set of  elements
 \be \mathcal J_\alpha(\bs n) = \{  \tau_0 \in \mathcal C(\bs n) : \text{there exists a table $\bs u$ such that $\tau(\bs u)=\tau_0$ and } p(\bs u, \bs n) \ge \alpha \} \ee
 is an interval equal to $\mathcal I_\alpha(\bs n)$, so that (by \Cref{l:containsobserved})
\be
\big[ T(\bs n ), C_+ \big] \cap  \mathcal J_\alpha(\bs n) = \big[ T(\bs n), U_\alpha(\bs n) \big].
\ee
Then it suffices to show the following two claims. First, that the $f$ given in \Cref{a:efficient} satisfies $f(x) = 0$ if $x/n \in \mathcal J_\alpha$ and $f(x)=1$ otherwise. Second, that $f(x)$ can be computed in at most $(n+1)$ permutation tests for every $x\neq0$, and $2(n+1)$ permutation tests when $x=0$. In this case, \Cref{l:binarylog} shows that the binary search indicated in \Cref{a:efficient} will return $U_\alpha(\bs n)$ after performing at most $(n+1)( \lfloor \log_2 (n +1) + 2 \rfloor +1)$ permutation tests, as desired.

We begin with the claim that $f(x) = 0$ if $x/n \in \mathcal J_\alpha$ and $f(x)=1$ otherwise. Set $\tau_0=x/n$. If a vector $\bs v$ permutation tested in Step~\eqref{step5}  satisfies $p(\bs v, \bs n) \ge \alpha$, then  $\tau_0 = \tau(\bs v)$ lies in $\mathcal J_\alpha(\bs n)$, and $f$ correctly declares $\tau_0$ compatible. 

Otherwise, if $j=n+1$ and the last step declares $\tau_0$ incompatible, we must show that every possible potential outcome count vector $\bs u$ with $\tau(\bs u) = \tau_0$ satisfies $p(\bs u, \bs n) < \alpha$. To this end, fix an arbitrary $\bs u$ such that $\tau(\bs u) = \tau_0$. 
If $\bs u$ was tested for compatibility in Step~\eqref{step5}, then we must have $p(\bs u, \bs n) < \alpha$, since the permutation test of $\bs u$ must have rejected for $f$ to have declared $\tau_0$ incompatible. 

Therefore, we may suppose $\bs u$ was not tested. Let $j = u_{11} + u_{01}$, and let $v_{01}$ be the value determined in Step \eqref{stepb}. Let $\bs v$ be the vector determined by \eqref{stepa} for this value of $j$, using this value of $v_{01}$. If $\tau_0$ was tested declared incompatible by $f$, then $\bs v$ was permutation tested and rejected. Hence $\bs u \neq \bs v$ and $p(\bs v, \bs n) < \alpha$. Suppose $v_{01} + v_{10} \ge 1$. 
Using this
inequality and \Cref{l:monotonic}, we deduce that
\be\label{vinf}
p(\bs u, \bs n )  \le p(\bs v, \bs n) < \alpha,
\ee
since $\bs v = \bs u + k (1 , -1, - 1, 1)$ for some $k \in \mathbb{Z}_{>0}$, as desired.
Now suppose that $v_{10} = v_{01} = 0$. 
Let $\bs v'$ be the vector constructed from $v_{10}'=1$ using \eqref{constructv} and $j = u_{11} + u_{01}$. 
We claim that $\bs v'$ is possible, and hence was tested and rejected (in the second part of \eqref{step5}). Indeed, if $\bs v'$ is not possible given the observed data, then by \Cref{l:possibleinequalities} we find that $\bs u \neq \bs v$ is not possible, since since the set of possible coordinates in \eqref{interval} is an interval. Then $p(\bs  v', \bs n) < \alpha$, and as before we have $p(\bs u, \bs n )  \le p(\bs v', \bs n) < \alpha$.

Hence, $\bs u$ is rejected if $\bs v$ is, and we have shown that evaluating $f$ provides a valid method for checking the compatibility of $\tau_0$.
Finally, the claim that $f$ can be evaluated using $(n+1)$ permutation tests for $x\neq 0$, and $2(n+1)$ tests when $x=0$, follows from the fact that examining a given $j\in \unn{0}{n}$ requires at most one permutation test when $x\neq 0$ and at most two tests otherwise, as described in Step~\eqref{step5}. This completes the proof. 
\end{proof}

\subsection{Proof of Lemma~\ref{l:monotonic}}\label{s:monotonic}

We first provide some intuition for the proof of \Cref{l:monotonic}. 
We will see that the sampling distribution for the Neyman estimator $T(\bs w)$ of a generic table $\bs w$ is SD on $ m^{-1} \mathbb{Z}$. 
If all tables $\bs w$ had this property, the proof of \Cref{l:monotonic} would be fairly straightforward (by applying \Cref{l:monotonic2} below). However, there are exceptional tables that do not, due to parity issues.
Whenever we have a table of the form $\bs v = (a,0,0,b)$ with $a+b=n=2m$, the estimator $T(\bs w, \bs Z)$ is supported on $ m^{-1} (2\mathbb{Z})$ or $ m^{-1} (2\mathbb{Z}+1)$ instead of $m^{-1} \mathbb{Z}$. 
Most of our effort goes into showing that this is the only such bad case.






\begin{lem}\label{l:monotonic2} 
Fix observed data $\bs n$, and a count vector 
\be
\bs v = (v_{11}, v_{10}, v_{01}, v_{00}).
\ee
Suppose $\min(v_{10}, v_{01}) \ge 1$, and define
\be
\bs v' = (v_{11}+1, v_{10}-1, v_{01}-1, v_{00}+1).
\ee
Set 
\be
\bs v^\circ = (v_{11}, v_{10}-1, v_{01}-1, v_{00}),
\ee
and suppose that the pmf of $T(\bs v^\circ, \bs Z)$ is SD on the lattice $(m-1)^{-1} \mathbb{Z}$, where here  $\bs Z$ is uniformly distributed over $\mathcal Z(n-2)$. 
Then 
\be\label{lemclaim}
p( \bs v', \bs n) \ge p(\bs v, \bs n).
\ee
\end{lem}
\begin{proof}
We couple the distributions of $T(\bs v, \bs Z)$ and $T(\bs v', \bs Z)$ in the following way. Let $\bs w$ be a potential outcome table with count vector $\bs v$, and label the subjects so that $\bs w_1 = (1,0)$ and $\bs w_2 = (0,1)$. Let $\bs w'$ be a potential outcome table such that $\bs w'_1 = (1,1)$ and $\bs w'_2 = (0,0)$, and $\bs w'_i = \bs w_i$ for $i \ge 3$. 
Then $\bs w'$ has the count vector $\bs v'$.
Let $\bs {\tilde Y}$ be the observed outcome vector for $\bs w$, let $\bs {\tilde Y}'$ be the observed outcome vector for $\bs w'$, and 
recall from  \eqref{pvalue} that
\begin{align}\label{orange}
p(\bs w, \bs Y, \bs Z) = \P\left( \big| T(\bs{\tilde Y}, \bs{ \tilde Z} )  - \tau_0  \big| \ge \big| T(\bs Y, \bs Z )  - \tau_0  \big| \right).
\end{align}
by definition.

We may suppose $T(\bs Y, \bs Z )  - \tau_0 \neq 0$, otherwise $p( \bs v', \bs n) = p(\bs v, \bs n) = 1$ and the claim \eqref{lemclaim} is trivially true.
Then \eqref{orange} implies
\begin{align}
p(\bs w, \bs Y, \bs Z) &=
 \P\left( T(\bs{\tilde Y}, \bs{ \tilde Z} )  - \tau_0   \ge \big| T(\bs Y, \bs Z )  - \tau_0  \big| \right) \label{p1} \\
& +
 \P\left(  T(\bs{\tilde Y}, \bs{ \tilde Z} )  - \tau_0   \le  - \big| T(\bs Y, \bs Z )  - \tau_0  \big| \right). \label{p2}
\end{align}
To complete the proof, it suffices to show that each of the probabilities \eqref{p1} and \eqref{p2} increases when $\bs w$ is replaced $\bs w'$. Observe that 
\be
\E \big[ T(\bs{\tilde Y}, \bs{ \tilde Z} ) \big] = \tau(\bs w) = \tau_0,
\ee
and that  $T(\bs{\tilde Y}, \bs{ \tilde Z} )$ is symmetric about its mean, by \Cref{l:symmetric}. Then by symmetry, it suffices to show that \eqref{p1} increases when $\bs w$ is replaced with $\bs w'$. 

As a preliminary observation, note that $\tau_0 \in \mathcal{C}$ takes on values in the lattice $n^{-1} \mathbb Z$, while $T$ takes on values in $m^{-1} \mathbb Z= 2n^{-1}\mathbb Z$. However, we are considering the probability
\be
\P \left( T(\bs{\tilde Y}, \bs{ \tilde Z} )   \ge \big| T(\bs Y, \bs Z )  - \tau_0  \big| + \tau_0 \right),
\ee
from \eqref{p1}, and one checks that $\big| T(\bs Y, \bs Z )  - \tau_0  \big| + \tau_0$ is an element of $m^{-1} \mathbb{Z}$ regardless of the value of $\tau_0$, since $T\in m^{-1} \mathbb{Z}$. 

Now, given $T(\bs{\tilde Y} , \bs {\tilde Z})$ for some fixed realization of $\bs {\tilde Z}$, we consider what happens to its value as $\bs w$ changes to $\bs w'$, so that $\bs{\tilde Y}$ is replaced by $\bs{\tilde Y}'$. There are three cases.
\begin{enumerate}
\item $\tilde Z_1 = \tilde Z_2$. Then the first two subjects are in the same group. The change $(1,0) \mapsto (1,1)$ and $(0,1) \mapsto (0,0)$ leaves $T$ invariant in this case.
\item $\tilde Z_1 = 1$ and $\tilde Z_2 =0$. The change of $(1,0) \mapsto (1,1)$ for the first subject does not affect $T$. The change $(0,1) \mapsto (0,0)$ for the second subject increases $T$ by $m^{-1}$. 
\item $\tilde Z_1=0$ and $\tilde Z_2 = 1$. The change of $(1,0) \mapsto (1,1)$ for the first subject decreases $T$ by $m^{-1}$. The change $(0,1) \mapsto (0,0)$ for the second subject does not affect $T$. 
\end{enumerate}
In the conclusion, the value of $T$ changes by $0$ or $\pm m^{-1}$. 
Then for $a\in \mathbb{Z}$, we have 
\begin{align}\label{couplingexp}
\P \left( T(\bs{\tilde Y'}, \bs{ \tilde Z} )   = \frac{a}{m}
 \right)
&= 
\P \left( T(\bs{\tilde Y}, \bs{ \tilde Z} )   = \frac{a-1}{m} 
\text{ and $(\tilde Z_1, \tilde Z_2) = (1, 0)$}
\right) \\
&+\P \left( T(\bs{\tilde Y}, \bs{ \tilde Z} )   = \frac{a}{m} \text{ and $\tilde Z_1 = \tilde Z_2$} \right) \\
&+\P \left( T(\bs{\tilde Y}, \bs{ \tilde Z} )   = \frac{a+1}{m} 
\text{ and $(\tilde Z_1, \tilde Z_2) = (0, 1)$}
\right).
\end{align}

As noted below \eqref{p2}, it suffices to show that 
\be\label{pineq}
\P \left( T(\bs{\tilde Y'}, \bs{ \tilde Z} )   \ge \big| T(\bs Y, \bs Z )  - \tau_0  \big| + \tau_0 \right) \ge
\P \left( T(\bs{\tilde Y}, \bs{ \tilde Z} )   \ge \big| T(\bs Y, \bs Z )  - \tau_0  \big| + \tau_0 \right) ,
\ee
since this implies that \eqref{p1} increases if $\bs v$ is replaced by $\bs v'$. Summing \eqref{couplingexp}, we get 
\begin{align}
\P &\left( T(\bs{\tilde Y'}, \bs{ \tilde Z} )   \ge \big| T(\bs Y, \bs Z )  - \tau_0  \big| + \tau_0 \right)\\
&= \P \left( T(\bs{\tilde Y}, \bs{ \tilde Z} )   \ge \big| T(\bs Y, \bs Z )  - \tau_0  \big| + \tau_0  + m^{-1} \right)\\
&+ \P \left( T(\bs{\tilde Y}, \bs{ \tilde Z} )   = \big| T(\bs Y, \bs Z )  - \tau_0  \big| + \tau_0 \text{ and $\tilde Z_1 = \tilde Z_2$}  \right) \\
&+ \P \left( T(\bs{\tilde Y}, \bs{ \tilde Z} )   =  \big| T(\bs Y, \bs Z )  - \tau_0  \big| + \tau_0  \text{ and $(\tilde Z_1, \tilde Z_2)=(1,0)$}  \right)\\
&+ \P \left( T(\bs{\tilde Y}, \bs{ \tilde Z} )   =  \big| T(\bs Y, \bs Z )  - \tau_0  \big| + \tau_0 - m^{-1} \text{ and $(\tilde Z_1, \tilde Z_2)=(1,0)$}  \right).
\end{align}
Using the previous equality, we see that \eqref{pineq} is equivalent to the statement that 
\begin{align}
& \P \left( T(\bs{\tilde Y}, \bs{ \tilde Z} )   = \big| T(\bs Y, \bs Z )  - \tau_0  \big| + \tau_0 \text{ and $\tilde Z_1 = \tilde Z_2$}  \right) \\
&+ \P \left( T(\bs{\tilde Y}, \bs{ \tilde Z} )   =  \big| T(\bs Y, \bs Z )  - \tau_0  \big| + \tau_0  \text{ and $(\tilde Z_1, \tilde Z_2)=(1,0)$}  \right)\\
&+ \P \left( T(\bs{\tilde Y}, \bs{ \tilde Z} )   =  \big| T(\bs Y, \bs Z )  - \tau_0  \big| + \tau_0 - m^{-1} \text{ and $(\tilde Z_1, \tilde Z_2)=(1,0)$}  \right)\\
&\ge \P \left( T(\bs{\tilde Y}, \bs{ \tilde Z} )   =  \big| T(\bs Y, \bs Z )  - \tau_0  \big| + \tau_0  \right),
\end{align}
which rearranges to 
\begin{align}\label{prev0}
&\P \left( T(\bs{\tilde Y}, \bs{ \tilde Z} )   =  \big| T(\bs Y, \bs Z )  - \tau_0  \big| - m^{-1} + \tau_0  \text{ and $(\tilde Z_1, \tilde Z_2)=(1,0)$}  \right)\\
&\ge \P \left( T(\bs{\tilde Y}, \bs{ \tilde Z} )   =  \big| T(\bs Y, \bs Z )  - \tau_0  \big| + \tau_0  \text{ and $(\tilde Z_1, \tilde Z_2)=(0,1)$}  \right).
\end{align}
Let $\bs {\tilde Z}^\circ = (\tilde Z_3^\circ, \dots \tilde Z_n^\circ)$ be uniformly distributed on  $\mathcal Z(n-2)$ (recall \eqref{zn}),  let $\bs w^{\circ} = (\bs w_3 , \dots, \bs w_n)$, and let ${\bs {\tilde Y}}^\circ$ be the vector of observed outcomes corresponding to $\bs {\tilde Z}^\circ$ and $\bs w^{\circ}$. Observe that $\bs v^\circ$ is the count vector for $\bs w^\circ$. Then \eqref{prev0} may be rewritten as 
\begin{align}\label{prev02}
&\P \left( T(\bs{\tilde Y}, \bs{ \tilde Z} )   =  \big| T(\bs Y, \bs Z )  - \tau_0  \big| - m^{-1} + \tau_0  \;\Big|\; (\tilde Z_1, \tilde Z_2)=(1,0)  \right)\P \big( (\tilde Z_1, \tilde Z_2)=(1,0)  \big)\\
&\ge \P \left( T(\bs{\tilde Y}, \bs{ \tilde Z} )   =  \big| T(\bs Y, \bs Z )  - \tau_0  \big| + \tau_0  \;\Big|\; (\tilde Z_1, \tilde Z_2)=(0,1)  \right)
\P\big( (\tilde Z_1, \tilde Z_2)=(0,1) \big),
\end{align}
and further as
\begin{align}\label{prev1}
&\P \left(  T(\bs{\tilde Y}^\circ, \bs{ \tilde Z}^\circ )   =  A - (m-1)^{-1} \right)\ge \P \left(  T(\bs{\tilde Y}^\circ, \bs{ \tilde Z}^\circ )   =  A  \right),
\end{align}
where 
\be
A = \frac{m}{m-1} \cdot  \Big(\big| T(\bs Y, \bs Z )  - \tau_0  \big| + \tau_0\Big) \in  (m-1)^{-1} \mathbb{Z}.
\ee
In the previous computations, we used the fact that the net contribution to $T(\bs{\tilde Y}, \bs{ \tilde Z} )$ from the first two subjects is zero, since we assumed that $\bs w_1 = (1,0)$ and $\bs w_2 = (0,1)$. 
We also used that the conditional distribution of $(\tilde Z_3, \dots \tilde Z_n)$, conditional on either $(\tilde Z_1, \tilde Z_2)=(1,0)$ or
 $(\tilde Z_1, \tilde Z_2)=(0,1)$, is uniform on the set  $\mathcal Z(n-2)$, and that the probabilities of these two events are equal.
Observe that
\be\label{rightofmean}
A  > \frac{m}{m-1} \cdot \tau_0  = \E\Big[ T\big(\bs{\tilde Y}^\circ, \bs{ \tilde Z}^\circ \big) \Big],
\ee 
since we assumed earlier that $T(\bs Y, \bs Z )  - \tau_0 \neq 0$.
By hypothesis, the pmf  of $T\big(\bs{\tilde Y}^\circ, \bs{ \tilde Z}^\circ\big)$ is SD, since it corresponds to the count vector $\bs v^\circ$. This fact, and the fact that $A$ is strictly greater than the mean by \eqref{rightofmean}, together imply that   \eqref{prev1} holds. Since \eqref{prev1} is equivalent to \eqref{pineq}, we see that \eqref{pineq} holds, which completes the proof.
\end{proof}

We next prove some lemmas to set up an induction.

\begin{lem}\label{l:sumwith01}
Let $X_1$ be SD on $m^{-1} \mathbb{Z}$, and let $X_2$ be uniformly distributed on $\{ 0 , m^{-1}\}$. Then $X_1+X_2$ is SD on $m^{-1}\mathbb{Z}$.
\end{lem}
\begin{proof}
It suffices to show that $m(X_1  +X_2)$ is SD on $\mathbb{Z}$. Denote the pdf of $mX_1$ by $f_1$. Then the pdf $g$ of  $m(X_1  +X_2)$ is given by $g(x) = \frac{1}{2}\big(f_1(x) + f_1(x-1)\big)$. Suppose that $mX_1$ is symmetric about $a \in \mathbb{Z}$. Then it is clear that $g(x)$ is symmetric about $a+\frac{1}{2}$. To show that $g(x)$ is decreasing, it suffices to show that for any $b\ge a + 1/2$, we have $g(b) \ge g(b+1)$. 
We find
\be
g(b) = \frac{1}{2}\big(f_1(b) + f_1(b-1)\big)
\ge  \frac{1}{2}\big(f_1(b+1) + f_1(b)\big) = g(b+1),
\ee
as desired. In the previous equation, we used $f_1(b) \ge f_1(b+1)$, which is true since $f_1$ is SD, and $f_1(b-1) \ge f_1(b)$. The latter is true by the SD property of $f_1$ if $b \ge a + 1$. If $b = a+ 1/2$, then this follows from the symmetric of $f_1$ about $a$.
\end{proof}

\begin{lem}\label{l:induct1}
Fix a potential outcome count vector 
\be
\bs v = (v_{11}, v_{10}, v_{01}, v_{00}),
\ee
and suppose at least one of the following conditions holds.
\begin{enumerate}
\item We have
$v_{10} \ge 2$, and the pmf for $T(\bs v- \bs a, \bs Z')$ is SD on $(m-1)^{-1} \mathbb{Z}$ for every choice of 
\be
\bs a \in 
\big\{ (1,1,0,0), (0,2,0,0) , (0,1,1,0), (0,1,0,1)
\big\} 
\ee
such that $\bs v- \bs a$  has nonnegative entries. Here, $\bs Z'$ denotes a random variable uniformly distributed on $\mathcal Z(n-2)$.
\item We have
$v_{01} \ge 2$, and the pmf for $T(\bs v- \bs a, \bs Z')$ is SD on $(m-1)^{-1} \mathbb{Z}$ for every choice of 
\be
\bs a \in 
\big\{ (1,0,1,0), (0,1,1,0) , (0,0,2,0), (0,0,1,1)
\big\} 
\ee
such that $\bs v- \bs a$  has nonnegative entries. 
\end{enumerate}
Then the pmf of $T(\bs v, \bs Z)$ is SD on $m^{-1} \mathbb{Z}$.
\end{lem}
\begin{proof}
We prove the claim only for the case $v_{10} \ge 2$, since the proof in the case that $v_{01} \ge 2$ is similar. By relabeling, we may suppose that the first subject has the potential outcome vector $\bs w_1 = (1,0)$. We sample $\bs Z$ in the following way. First, choose subject $j$ from $\{2, \dots ,n \}$ uniformly at random and consider the pair $(\bs w_1, \bs w_j)$. Then sample $(Z_1, Z_j)$ uniformly at random from the assignments $(Z_1 =1, Z_j = 0)$ and 
$(Z_1 =0, Z_j = 1)$. Finally, assign the remaining $n-2$ subjects to groups by an independent randomization to equal groups (chosen uniformly at random). 
We will show that conditional distribution of $T(\bs v, \bs Z)$, conditional on the choice of $j$, is SD with a mean of $\E\big[T(\bs v, \bs Z)\big]$. This suffices to prove the theorem, since the unconditional pmf for $T(\bs v, \bs Z)$ is the weighted sum of such conditional pmfs. There are four cases.
\begin{enumerate}
\item $\bs w_j = (1,1)$. Then $w_1(1) - w_j(0) = 0$, and 
$w_j(1) - w_1(0) = 1$, so the distribution of 
\be\label{width}
Z_1 w_1(1) + (1-Z_1)w_1(0) + Z_j w_j(1) + (1-Z_j)w_j(0) 
\ee
is uniform on the set $\{0,1\}$. Further, the contribution to $T$ from the other $n-2$ subjects, after the independent randomization is also symmetric decreasing on $m^{-1} \mathbb{Z}$, by assumption.  Then  the sum of these (conditionally) independent variables is SD with mean $\E\big[T(\bs v, \bs Z)\big]$, by \Cref{l:sumwith01}.

\item $\bs w_j = (1,0)$. One checks that the distribution \eqref{width} is constant and equal to $1$, and that the same reasoning as in the previous case applies, since the sum of a random variable that is SD and the constant $1$ is still SD.

\item $\bs w_j = (0,1)$. The distribution of \eqref{width} is constant and equal to $0$, and the same reasoning as in the previous point applies.
\item $\bs w_j = (0,0)$. The distribution of \eqref{width} is uniform on $\{0,1\}$, and we can apply \Cref{l:sumwith01}.
\end{enumerate}
This completes the proof.
\end{proof}

\begin{lem}\label{l:evens}
Fix any potential outcome count vector of the form
\be\label{aboveform}
\bs v = (v_{11}, 0 , 0, v_{00}).
\ee
Then the pmf of $T(\bs v, \bs Z)$ is SD on $m^{-1} (2\mathbb{Z})$ if $v_{11}$ is even. It is SD on $m^{-1} (2\mathbb{Z}+1)$ if $v_{11}$ is odd.
\end{lem} 
\begin{remark}
Note that this statement concerns the lattices  $m^{-1} (2\mathbb{Z})$  and $m^{-1} (2\mathbb{Z}+1)$, instead of the lattice $m^{-1}\mathbb{Z}$, due to the parity issue mentioned in the introduction to this subsection.
\end{remark}
\begin{proof}

We prove the claim for all even $n \in \mathbb{Z}_{\ge 0}$ by induction. The base case $n =0$ is trivial. For the induction step, we suppose that the claim holds for $n-2$ and will show it is true for $n$. Fix an arbitrary $\bs v$ of the form \eqref{aboveform} with $v_{11} + v_{00} =n$. We suppose that $v_{11} \ge 2$. When $v_{11} = 0$, the conclusion is trivial, and when $v_{11} = 1$, we can apply the reasoning below with $v_{00}$ instead of $v_{11}$. 

By relabeling, we may suppose $\bs w_1 = (1,1)$. We sample $\bs Z$ as in the previous proof by choosing a partner $\bs w_j$ for $\bs w_1$ uniformly at random, randomizing the pair $\bs w_1$ and $\bs w_j$ to treatment and control, and independently randomizing the other $n-2$ subjects to equal groups. 

We consider first the case that $\bs w_2 = (1,1)$. Then the distribution of
\be\label{width2}
Z_1 w_1(1) + (1-Z_1)w_1(0) + Z_j w_j(1) + (1-Z_j)w_j(0) 
\ee
conditional on the choice of partner and that fact that subjects $1$ and $2$ are assigned to different groups is constant and equal to $0$. This implies we are done by induction, since $v_{11} -2$, the number of $(1,1)$ potential outcome vectors in the remaining $n-2$ subjects, has the same parity as $v_{11}$, and the sampling distributions of 
\be\label{supp1}
\sum_{i=1}^n Z_i w_i(1) + (1-Z_i)w_i(0)
\ee
and
\be\label{supp2}
\sum_{i=3}^n Z_i w_i(1) + (1-Z_i)w_i(0)
\ee
are then both supported and SD on either  $m^{-1} (2\mathbb{Z})$  or $m^{-1} (2\mathbb{Z}+1)$, as desired. 

We next consider the case where $\bs w_2  = (0,0)$. Then the conditional distribution of \eqref{width2} is uniform on $\{-1, 1\}$. We are again done by induction, since there are $v_{11}-1$ potential outcome vectors of the form $(1,1)$ in the remaining $n-2$ subjects, and this number has the opposite parity to $v_{11}$. Reasoning as in the proof of \Cref{l:sumwith01}, we see that the sums \eqref{supp1} and \eqref{supp2} are such that one is supported and SD on $m^{-1} (2\mathbb{Z})$  and one is supported and SD on $m^{-1} (2\mathbb{Z}+1)$, as desired.

%
Combining the conclusions of these two cases completes the proof.
\end{proof}

\begin{lem}\label{l:oneodd}
Let $n  = 2m$ for some $m \in \mathbb{Z}_{>0}$.
Fix a potential outcome count vector 
\be
\bs v = (v_{11}, 1 , 0, v_{00}) \text{ or } (v_{11}, 0 , 1, v_{00}).
\ee
Then the pmf of $T(\bs v, \bs Z)$ is SD on $m^{-1} \mathbb{Z}$.
\end{lem}
\begin{proof}
We consider only the first case, where $v_{10} = 1$ and $v_{01} = 0$, since the argument for the other case is similar. 
By relabeling, we may suppose $\bs w_1 = (1,0)$, and we resample $\bs Z$ as in the proof of \Cref{l:induct1} by first choosing uniformly at random a partner $\bs w_j$ for $\bs w_1$ from the set $\{ \bs w_2, \dots, \bs w_n \}$, sampling $(Z_1, Z_j)$ uniformly at random from the assignments $(Z_1 =1, Z_j = 0)$ and 
$(Z_1 =0, Z_j = 1)$, and assigning the remaining $n-2$ subjects to groups by an independent, uniform randomization to equal groups.
Let $\bs v^\circ$ denote the (random) count vector for the potential outcome table given by $\{ \bs w_2, \dots, \bs w_n \} \setminus \{\bs w_j\}$, and let $\bs Z'$ denote a vector uniformly distributed on $\mathcal Z(n-2)$. 

We now condition on the choice of $\bs w_j$. By \Cref{l:evens}, the conditional distribution of $ T( \bs v^\circ, \bs Z')$ for the remaining $n-2$ people is SD on $(m-1)^{-1} (2 \mathbb{Z})$ or $(m-1)^{-1}(2\mathbb{Z} + 1$). Similarly, the conditional distribution of \eqref{width} for the partnership is always uniform on $\{0,1\}$, since $\bs w_j = (0,0)$ or  $\bs w_j = (1,1)$. 
Then the conditional pmf for $T(\bs v, \bs Z)$ is SD on $ m^{-1}\mathbb{Z}$ by \Cref{l:sumwith01}, with mean independent of the choice of $\bs w_j$ and equal to the unconditional mean $\E\big[ T(\bs v, \bs Z) \big]$. Since this holds for every choice of $\bs w_j$, it holds for the unconditional distribution of $T(\bs v, \bs Z)$, as desired.
\end{proof}
The proof of the following lemma is somewhat computational, so we defer it to the next subsection. 
\begin{lem}\label{l:bigcomputation}
Let $n  = 2m$ for some $m \in \mathbb{Z}_{>0}$.
Fix a potential outcome count vector 
\be
\bs v = (v_{11}, 2 , 0, v_{00}) .\ee
Then the pmf of $T(\bs v, \bs Z)$ is SD on $m^{-1} \mathbb{Z}$.
\end{lem}

\begin{cor}\label{c:bigcomputation}
Let $n  = 2m$ for some $m \in \mathbb{Z}_{>0}$. 
Fix a potential outcome count vector 
\be
\bs v = (v_{11}, 1 , 1, v_{00}) \text{ or } (v_{11}, 0 , 2, v_{00}).
\ee
Then the pmf of $T(\bs v, \bs Z)$ is SD on $m^{-1} \mathbb{Z}$.
\end{cor}
\begin{proof}
The pmfs of $T(\bs v, \bs Z)$ for the given vectors are translates  of the pmf corresponding to $T(\bs v, \bs Z)$ for
\be
\bs v = (v_{11}, 2, 0, v_{00}),
\ee
so we are done by the previous lemma.
\end{proof}

\begin{lem}\label{l:final}
Let $n  = 2m$ for some $m \in \mathbb{Z}_{>0}$.
Fix any potential outcome count vector 
\be\label{finalform}
\bs v = (v_{11}, v_{10}, v_{01}, v_{00}),
\ee
and suppose $v_{10} + v_{01} \ge 1$. Then the pmf of $T(\bs v, \bs Z)$ is SD on $m^{-1} \mathbb{Z}$.
\end{lem}
\begin{proof}
We induct on $n$. The base case $n = 0$ is trivial. For the induction step, suppose that the claim is true for $n-2$, and fix any $\bs v$ corresponding to a potential outcome table with $n$ subjects such that $\min( v_{10}, v_{01} ) \ge 1$. If $(v_{10}, v_{01}) = (1,0)$ or $(v_{10}, v_{01}) = (0,1)$, we are done by \Cref{l:oneodd}. If $v_{01} + v_{01} = 2$, we are done by \Cref{l:bigcomputation} and \Cref{c:bigcomputation}. 

In the case $v_{01} + v_{01} > 2$, we will use \Cref{l:induct1} to show the claim is true for the given $\bs v$. In this case, suppose first that $v_{10} \ge 2$ and $v_{01} \ge 1$. Then by the induction hypothesis, \Cref{l:oneodd}, \Cref{l:bigcomputation}, and \Cref{c:bigcomputation}, the first condition given in the statement of \Cref{l:induct1} holds. The case where $v_{10} \ge 1$ and $v_{01} \ge 2$ is analogous. This completes the proof.
\end{proof}

\begin{proof}[Proof of \Cref{l:monotonic}]
This is an immediate consequence of combining \Cref{l:final} and \Cref{l:monotonic2}.
\end{proof}

\subsection{Proof of Lemma~\ref{l:bigcomputation}}
We first recall a formula of Copas \cite{copas1973randomization}. Given a potential outcome table $\bs w$ and a randomization $\bs Z$, we define a \emph{treatment count vector} $\bs x$ by 
\be
\bs x = (x_{11}, x_{10}, x_{01}, x_{00}), \qquad x_{ab} = \sum_{j=1}^n Z_j \one\{\bs w_j = (a,b) \}.
\ee
Note that
\be
x_{11} + x_{10} +  x_{01} +  x_{00} = m,
\ee
if $n=2m$ and we randomize into equal groups.

\begin{lem}[\cite{copas1973randomization}]\label{l:copas}
Fix a potential outcome count vector $\bs v$. For any $s_0, s_1 \in \mathbb{Z}_{\ge 0}$, the probability of observing a treatment count vector $\bs x$ such that 
\be
s_1 = x_{11} + x_{10}, \qquad s_0  = (v_{11} - x_{11}) + (v_{01} - x_{01} )
\ee
is
\be\label{pmfp}
p(s_1, s_0) = C_{\bs v} \sum_{x=-\infty}^\infty \binom{v_{11}}{x} \binom{v_{10}}{s_1 -x}
\binom{v_{01} }{v_{11} + v_{01} - s_0 -x} 
\binom{v_{00} }{m - v_{11} - s_1 - v_{01} +s_0 +x},
\ee
where $C_{\bs v} > 0$ is a constant depending only on $\bs v$. 
\end{lem}

\begin{proof}[Proof of \Cref{l:bigcomputation}]
We explicitly compute  the distribution of $T(\bs v, \bs Z)$. We first note that direct computation shows that $m \E \big[T (\bs v, \bs Z)\big] =1$, and we know that the distribution of $ T(\bs v, \bs Z)$ is symmetric about its mean by \Cref{l:symmetric}. 

 Adopt the notation of \Cref{l:copas}. Then  
\be\label{subs}
s_1 = x_{11} + x_{10}, \qquad s_0  = v_{11} - x_{11}.
\ee
With $p$ defined as in \eqref{pmfp}, and using the assumed form of $\bs v$, we have 
\begin{align}
p(s_1, s_0) &= C_{\bs v} \sum_{x} \binom{v_{11}}{x} \binom{2}{s_1 -x}
\binom{0 }{v_{11}  - s_0 -x} 
\binom{v_{00} }{m - v_{11} - s_1 - v_{01} +s_0 +x}\\
&=
C_{\bs v}  \sum_{x} \binom{v_{11}}{x} \binom{2}{x_{11} + x_{10} -x}
\binom{0 }{x_{11} -x} 
\binom{v_{00} }{m - v_{11} - s_1  +s_0 +x}.
\end{align}
For this to be nonzero, we must have $x=x_{11}$. Using this and \eqref{subs}, we get
\be\label{pepper}
p(s_1, s_0) = p(x_{11}, x_{10}) = C_{\bs v} \binom{v_{11}}{x_{11}} 
\binom{2}{x_{10}}
\binom{v_{00} }{m    -x_{11}  -x_{10}}.
\ee
We abbreviate $j = mT(\bs v, \bs Z) -1 $ (centering by subtracting the expectation). Then (using $v_{01} = 0$)
\be
j = s_1 - s_0 - 1= x_{11} + x_{10} - v_{11} + x_{11} -1 =  2x_{11} 
+ x_{10} -v_{11} -1
.\ee
This yields
\be\label{j1}
x_{11} = \frac{j + v_{11} - x_{10}+1}{2}.
\ee
We also have
\be\label{j2}
\frac{v_{00} + v_{11}}{2} + 1 = m.
\ee
Using \eqref{j1} and \eqref{j2} in \eqref{pepper}, we get
\be\label{pepper2}
p (x_{11}, x_{10}) = C_{\bs v}
\binom{2}{x_{10}}
\binom{v_{11}}{ \frac{v_{11}}{2} + \frac{ 1 - x_{10}+j}{2}} 
\binom{v_{00} }{\frac{v_{00}}{2} + \frac{ 1- x_{10} -j  }{2}}.
\ee
We now consider two cases, depending on the parity of $v_{11}$. 

\paragraph{Case I: $v_{11}$ is even.} Then $v_{00}$ is also even, since $n$ is even. There are two subcases, depending on the parity of $j$. Suppose that $j$ is even. Then parity considerations from \eqref{j1} force $x_{10}=1$. Then the probability mass function becomes
\be
p(j) = 2 C_{\bs v}
\binom{v_{11}}{ \frac{v_{11}}{2} + \frac{ j}{2}} 
\binom{v_{00} }{\frac{v_{00}}{2} - \frac{ j  }{2}}.
\ee
When $j$ is odd, both $x_{10}=0$ and $x_{10} = 2$ are possible, and the pmf is 
\be
p(j) = 
C_{\bs v} \binom{v_{11}}{ \frac{v_{11}}{2} + \frac{ 1 + j}{2}} 
\binom{v_{00} }{\frac{v_{00}}{2} + \frac{ 1 -  j  }{2}}
+
C_{\bs v} \binom{v_{11}}{ \frac{v_{11}}{2} + \frac{ j - 1}{2}} 
\binom{v_{00} }{\frac{v_{00}}{2} + \frac{ - 1 - j }{2}}.
\ee 
Now it is a straightforward computation to show that $p(j)$ is decreasing for $j \ge 0$. We will show that $p(j) \ge p(j+1)$ for all $j \ge 0$. First, suppose that $j$ is even. Then, dividing through by $C_{\bs v}$. we must show that 
\be
2 
\binom{v_{11}}{ \frac{v_{11}}{2} + \frac{ j}{2}} 
\binom{v_{00} }{\frac{v_{00}}{2} - \frac{ j  }{2}}
\ge 
 \binom{v_{11}}{ \frac{v_{11}}{2} + \frac{ j + 2 }{2}} 
\binom{v_{00} }{\frac{v_{00}}{2} - \frac{ j }{2}}
+
 \binom{v_{11}}{ \frac{v_{11}}{2} + \frac{ j}{2}} 
\binom{v_{00} }{\frac{v_{00}}{2} + \frac{ - 2 - j }{2}}.
\ee
This is clear, since $\binom{n}{k}$ is symmetric and decreasing away from $ k = n/2$. 

Next, we suppose that $j$ is odd. We want to show
\be
 \binom{v_{11}}{ \frac{v_{11}}{2} + \frac{ 1 + j}{2}} 
\binom{v_{00} }{\frac{v_{00}}{2} + \frac{ 1 -  j  }{2}}
+
\binom{v_{11}}{ \frac{v_{11}}{2} + \frac{ j - 1}{2}} 
\binom{v_{00} }{\frac{v_{00}}{2} + \frac{ - 1 - j }{2}}
\ge 
2 \binom{v_{11}}{ \frac{v_{11}}{2} + \frac{ j+1}{2}} 
\binom{v_{00} }{\frac{v_{00}}{2} - \frac{ j  + 1 }{2}}.
\ee
This is again clear, for the same reason.


\paragraph{Case II: $v_{11}$ is odd.} Then $v_{11}$ is odd, since $n$ is even. There are two subcases, depending on the parity of $j$. Suppose that $j$ is odd.  Then parity considerations from \eqref{j1} force $x_{10}=1$. 
Then the probability mass function becomes
\be
p(j) = 2 C_{\bs v}
\binom{v_{11}}{ \frac{v_{11}}{2} + \frac{ j}{2}} 
\binom{v_{00} }{\frac{v_{00}}{2} - \frac{ j  }{2}}.
\ee
When $j$ is even, both $x_{10}=0$ and $x_{10} = 2$ are possible, and the pmf is 
\be
p(j) = 
C_{\bs v} \binom{v_{11}}{ \frac{v_{11}}{2} + \frac{ 1 + j}{2}} 
\binom{v_{00} }{\frac{v_{00}}{2} + \frac{ 1 -  j  }{2}}
+
C_{\bs v} \binom{v_{11}}{ \frac{v_{11}}{2} + \frac{ j - 1}{2}} 
\binom{v_{00} }{\frac{v_{00}}{2} + \frac{ - 1 - j }{2}}.
\ee 
The same verification as in the previous case shows that $p(j)$ is decreasing for $j \ge 0$.
\end{proof}

\subsection{Proof of Proposition~\ref{p:montecarlo}}\label{s:montecarloproof}
\begin{lem}\label{l:carrot}
Fix $\eps >0$ and $K \in \mathbb{Z}_{>0}$. Then
\be\label{oneside1}
\P\big(  p(\bs w, \bs Y, \bs Z )  - S  > \eps   \big) \le  \exp\left( - 2K \eps^2 \right).
\ee
and 
\be\label{oneside2}
\P\big( S - p(\bs w, \bs Y, \bs Z )   > \eps \big) \le  \exp\left( - 2K \eps^2 \right).
\ee
\end{lem}
\begin{proof}
We prove only the first claim, since the second is analogous. We have $\E[V_i ] = p(\bs w, \bs Y, \bs Z )$ and $ 0\le V_i \le 1$ by definition, so the one-sided version of Hoeffding's inequality yields 
\be
\P\left(  K \cdot p(\bs w, \bs Y, \bs Z ) -   \sum_{i=1}^K V_i   > K \eps \right)\le  \exp\left( - \frac{2 (K \eps)^2}{K} \right) =  \exp\left( - 2K \eps^2 \right),
\ee
as desired. 
\end{proof}

\begin{lem}\label{l:montecarlo}
Fix a potential outcome table $\bs y$ and $\alpha \in (0,1)$. 
\begin{enumerate}
\item 
With probability at least  $ 1 - 2 (n+1) \lfloor \log_2(n+1) + 3 \rfloor \exp(-2  K\eps^2)$, we have 
\begin{equation}\label{e:firstc}
\mathcal I_\alpha(\bs Y, \bs Z) \subset \mathcal I_{\alpha, \eps, K} (\bs Y, \bs Z).\end{equation}
\item With probability at least $ 1 - 2 (n+1) \lfloor \log_2(n+1) + 3 \rfloor \exp(- 2 K\eps^2)$, we have 
\begin{equation}\label{e:secondc}
\mathcal I_{\alpha, \eps, K}(\bs Y, \bs Z) \subset \mathcal I_{\alpha - 2\eps}(\bs Y, \bs Z).
\end{equation}
\end{enumerate}
\end{lem}
\begin{proof}
We begin with the first claim. 
The event where $\mathcal I_\alpha(\bs Y, \bs Z) \subset \mathcal I_{\alpha, \eps, K} (\bs Y, \bs Z)$ is contained in the event
\begin{equation} \label{e:twoevents}
\big\{ U_\alpha(\bs Y, \bs Z) \le U_{\alpha,\eps,K} (\bs Y, \bs Z) \big\} \cap  \big\{ L_\alpha(\bs Y, \bs Z) \ge L_{\alpha,\eps,K} (\bs Y, \bs Z)\big\}.
\end{equation}

We begin by bounding the probability that $U_\alpha(\bs Y, \bs Z) \le U_{\alpha,\eps,K} (\bs Y, \bs Z)$ does not hold. Let $N$ be the total number of permutation tests performed by \Cref{a:efficient} in the course of finding $U_\alpha(\bs Y, \bs Z)$. Letting $0$ represent acceptance and $1$ represent rejection, we let $\bs a = (a_i)_{i=1}^N \in \{0,1\}^N$ denote the sequences of acceptances and rejections made by the permutation tests in \Cref{a:efficient}. 
We condition on the observed data $(\bs Y, \bs Z)$, so that $\bs a$ becomes deterministic. 
Given this observed data, let $M$ denote the (random) number of approximate permutations performed by \Cref{a:approximate} before returning $U_{\alpha,\eps,K} (\bs Y, \bs Z)$. Let $\bs b = (b_i)_{i=1}^M \in \{0,1\}^M$ denote the corresponding (random) sequence of acceptances and rejections. 
Let $J$ be a random variable denoting the the smallest positive integer such that $a_J \neq b_J$. If $a_i = b_i$ for all $i\in \unn{1}{N}$, in which case $U_\alpha = U_{\alpha,\eps, K}$, we set $J = N+1$ and define $b_J = 0$. Note that $J$ is well defined, since it is not possible for \Cref{a:approximate} to terminate unless $\bs b$ differs from $\bs a$ at some location, or $M=N$ and $\bs b = \bs a$. Further, we have the deterministic bound $J \le N+1$.

The event $\big\{ U_\alpha(\bs Y, \bs Z) \le U_{\alpha,\eps,K} (\bs Y, \bs Z) \big\} $ is equal to the event that $b_J=0$. That is, in order that $U_\alpha \le U_{\alpha,\eps,K}$, at the first time the result of an approximate permutation test in \Cref{a:approximate} differs from the result of an exact permutation test, the approximate test must accept.
We therefore compute
\begin{align}
\P\Big(\big\{ U_\alpha(\bs Y, \bs Z) \le U_{\alpha,\eps,K} (\bs Y, \bs Z) \big\}^c\Big)
&= 
\P( b_J = 1)\notag \\
&=\sum_{i=1}^{N} \P( b_i =1 \text{ and } J=i)\notag\\
&\le \sum_{i=1}^{N} \P( b_i =1 \text{ and } a_i=0)\notag \\
&\le N \exp(-2 K\eps^2)\label{mango1}\\
& \le (n+1) \lfloor \log_2(n+1) + 3 \rfloor \exp(-2K\eps^2).\label{mango2}
\end{align}
To justify \eqref{mango1}, we note that $a_i=0$ implies $ p(\bs w, \bs Y, \bs Z) \ge \alpha$ for the table $\bs w$ considered at the $i$-th step of \Cref{a:efficient}. For \Cref{a:approximate} to reject $\bs w$, it must be rejected by \Cref{a:approximatetest}. In the notation of \Cref{a:approximate}, this happens when $S + \eps < \alpha$. Then by \eqref{oneside1}, \Cref{a:approximatetest} rejects $\bs w$ with probability at most $\exp(-2 K\eps^2)$. In \eqref{mango2}, we used the fact that at most 
\be (n+1) \lfloor \log_2(n+1) + 3 \rfloor
\ee permutation tests are required in \Cref{a:approximate} to find $U_\alpha$, which was shown in the proof of \Cref{t:main}.
The bound on the probability that $L_\alpha \ge L_{\alpha, \eps, K}$ fails to holds follows by a nearly identical argument. From a union bound over the complements of the two events in \eqref{e:twoevents}, we deduce \eqref{e:firstc}.

For the proof of the second claim, we keep the notation $\bs a$ and $\bs b$, but now define $\bs a$ using the interval $\mathcal I_{\alpha - 2\eps}(\bs Y, \bs Z)$. The event 
$\mathcal I_{\alpha, \eps, K}(\bs Y, \bs Z) \subset \mathcal I_{\alpha - 2\eps}(\bs Y, \bs Z)$
is contained in the event
\begin{equation} \label{e:twoevents2}
\big\{ U_{\alpha-2\eps}(\bs Y, \bs Z) \ge U_{\alpha,\eps,K} (\bs Y, \bs Z) \big\} \cap  \big\{ L_{\alpha-2\eps}(\bs Y, \bs Z) \le L_{\alpha,\eps,K} (\bs Y, \bs Z)\big\}.
\end{equation}
The event $\big\{ U_{\alpha-2\eps}(\bs Y, \bs Z) \ge U_{\alpha,\eps,K} (\bs Y, \bs Z) \big\}$ is equal to the event 
\begin{equation}
\{b_J = 1 \} \cup \{ J = N+1 \}. 
\end{equation}
We compute
\begin{align}
\P\Big(\big\{ U_{\alpha-2\eps}(\bs Y, \bs Z) \ge U_{\alpha,\eps,K} (\bs Y, \bs Z) \big\}^c\Big)
&= 
\P( b_J = 0 \text{ and } J \le N)\\
&=\sum_{i=1}^{N} \P( b_i =0 \text{ and } J=i)\\
&\le \sum_{i=1}^{N} \P( b_i =0 \text{ and } a_i=1)\\
&\le N \exp(-2K\eps^2)\label{mango3}\\
& \le (n+1) \lfloor \log_2(n+1) + 3 \rfloor\exp(-2K\eps^2).\label{mango4}
\end{align}
To justify \eqref{mango3}, we note that $a_i=1$ implies $ p(\bs w, \bs Y, \bs Z) < \alpha -2\eps$ for the table $\bs w$ considered at the $i$-th step of \Cref{a:efficient}. For \Cref{a:approximate} to accept $\bs w$, it must be accepted by \Cref{a:approximatetest}. That is, it must be the case that $S + \eps \ge \alpha$. Using the preceding bound on $p(\bs w, \bs Y, \bs Z)$, we  bound the acceptance probability as 
\begin{align}
\P(S \ge \alpha  - \eps ) & \le 
\P\big(S  + \alpha - 2\eps \ge \alpha  - \eps + p(\bs w, \bs Y, \bs Z)\big) \\
& = \P\big(S - p(\bs w, \bs Y, \bs Z)  \ge  \eps  \big) \\
& 
\le  \exp ( -2K \eps^2),
\end{align}
where the last step follows by \eqref{oneside2}. The estimate for the event $\big\{ L_{\alpha-2\eps}(\bs Y, \bs Z) \le L_{\alpha,\eps,K} (\bs Y, \bs Z)\big\}$ is identical, and is shown by essentially the same argument. This completes the proof of \eqref{e:secondc} after using a union bound.
\end{proof}

\begin{proof}[Proof of \Cref{p:montecarlo}]
For the first part, it suffices to choose $K$ large enough so that \be \P ( \mathcal I_{\alpha -\eps} \subset \mathcal I_{\alpha-\eps, \eps, K}) \ge 1 - \eps.\ee
Then the claim follows from \Cref{l:montecarlo} and the fact that
\begin{equation}
4 n \log_2(n) \ge 2(n+1) \lfloor \log_2(n+1) + 3 \rfloor
\end{equation}
for $n\ge 15$. 
The second claim is immediate from \Cref{l:montecarlo} and the previous inequality.
\end{proof}
\section{Unbalanced Trials}\label{a:unbalanced}
\begin{proof}[Proof of \Cref{t:unbalanced}]
We begin with the first claim. For each possible potential outcome count vector $\bs v$, a valid $p$-value for the null $\hat {\bs v} = \bs v$ is computed either by a direct application of \Cref{a:approximatetest2} in step 2(c), or by an alternative calculation in step 2(d), and  $\theta(\bs v)$ is included in $\mathcal I_{\alpha, K}$ if this $p$-value is greater than $\alpha$. (See \Cref{r:pexact}.)  Note that it is possible to have $\theta(\bs v_1) = \theta(\bs v_2)$ for distinct count vectors $\bs v_1$ and $\bs v_2$, for which $\bs v_1$ is rejected and $\bs v_2$ is accepted, so that $\theta(\bs v_1)$ is ultimately included in the final confidence interval despite the rejection of $\bs v_1$.
Then the value $\theta(\bs v)$ corresponding to the count vector $\bs v$ such that $\bs v = \bs {\hat v}$ will be included the returned interval $\mathcal I_{K, \alpha}$ with probability at least $1-\alpha$, by \eqref{e:validp}. This shows \eqref{unbalancedclaim1}.

We now turn to the second claim, about the time complexity. Steps 1 and 2 are repeated, as a group, at most $n+1$ times. Step 1 is constant time, and Step 2 is done at most $n+1$ times. Considering the components of step 2, we see that steps 2(a), 2(b), and 2(e) can be done in constant time. It remains to consider the contributions of steps 2(c) and 2(d). Unraveling these loops, we see that the final time complexity will be a factor of $n^2$ greater than the sum of the complexities of steps 2(c) and 2(d). 

Step 2(c) uses the Monte Carlo permutation test in \Cref{a:approximatetest2}. Using the Fisher--Yates algorithm, each permutation in \Cref{a:approximatetest2} can be generated in $O(n)$ time, and each $V_i$ can be calculated in $O(n)$ time. Since each Monte Carlo permutation test uses $K$ samples, we find that \Cref{a:approximatetest2} requires $O( K n)$ arithmetical operations.



We next consider the time required for step 2(d). The time complexity of computing $\{ \bs {q'}^{(i)} \}_{i=1}^K$ given $\{ \bs {q}^{(i)} \}_{i=1}^K$ is $O(K)$, since for each $i\in\unn{1}{K}$, $\bs {q}^{(i)}$ can be computed from $\bs {q'}^{(i)}$ in constant time.
Additionally, the time complexity of computing the approximate $p$-value \eqref{approxpval} given the $K$ vectors $\{ \bs q'^{(i)} \}_{i=1}^K$ is also $O(K)$. There at most $n$ points in the line segment  $\mathcal L$, so the total complexity of step 2(d) is therefore $O(Kn)$.

Collecting these observations, we find that the total time complexity of the algorithm is $O(Kn^3)$. 
\end{proof}

\section{Proof of Proposition~\ref{p:missing}}\label{s:missingproofs}

\begin{proof}[Proof of \Cref{p:missing}]
First, note that by \Cref{p:coverage}, 
\be
\P\big ( \tau(\bs y) \in \mathcal J_\alpha(\bs M, \bs Z) \big ) \ge 1 - \alpha,
\ee
where 
\be
\mathcal J_\alpha(\bs M, \bs Z) = \bigcup_{\bs Y' \in \mathcal Y(\bs M)} \mathcal I_\alpha ( \bs Y', \bs Z),
\ee
and $\mathcal Y(\bs M )$ consists of all vectors of realized outcomes $\bs Y'$ compatible with the partially observed vector $\bs M$. To complete the proof of the theorem it then suffices to show that
\be
\mathcal J_\alpha(\bs M, \bs Z) \subset \mathcal I^\circ_\alpha(\bs M, \bs Z),
\ee 
or equivalently,
\be
\mathcal I^\circ_\alpha(\bs M, \bs Z)^c 
\subset \mathcal J_\alpha(\bs M, \bs Z)^c = 
\bigcap_{\bs Y' \in \mathcal Y(\bs M)} \mathcal I_\alpha ( \bs Y', \bs Z)^c.
\ee
Fix some $\tau_0 \in \mathcal I^\circ_\alpha(\bs M, \bs Z)^c$. 
We may suppose that 
\be \tau_0 >  \max\big (U_\alpha(\bs Y^{(+)}, \bs Z), T(\bs Y^{(+)}, \bs Z) \big),
\ee
since the analogous argument in the complementary case is similar. 
Then we must show that all  tables $\bs w$ such that $\tau(\bs w) = \tau_0$ are 
incompatible with all choices of $( \bs Y', \bs Z)$ with $\bs Y' \in \mathcal Y(\bs M)$. Fix such a $\bs w$ and a choice of $\bs Y'$.

By the definition of $\bs Y^{(+)}$, and since $T(\bs Y^{(+)}, \bs Z)$ and $T (\bs Y', \bs Z)$ are coupled through $\bs Z$, we have 
\be
\tau_0 \ge T(\bs Y^{(+)}, \bs Z) \ge T (\bs Y', \bs Z).
\ee
Then 
\be
\P\left( \big| T(\bs{ \tilde Y}, \bs{ \tilde Z} )  - \tau_0  \big| \ge \big| T(\bs Y', \bs Z )  - \tau_0  \big| \right) 
\le 
\P\left( \big| T( \bs{ \tilde Y} , \bs{ \tilde Z} )  - \tau_0  \big| \ge \big| T(\bs Y^{(+)}, \bs Z )  - \tau_0  \big| \right) 
< \alpha,
\ee
so $\bs w$ is incompatible with $( \bs Y', \bs Z)$, as desired. 
\end{proof}

\end{document}